\documentclass[review]{elsarticle}

\usepackage{lineno,hyperref}

\modulolinenumbers[5]

\journal{Journal of \LaTeX\ Templates}

\begin{document}

\begin{frontmatter}

\title{Double-distribution-function discrete Boltzmann model for combustion}

\author[address1]{Chuandong Lin}

\author[address2,address3,address4]{Aiguo Xu\fnref{myfootnote1}}
\fntext[myfootnote1]{Corresponding author. Email addresses: Xu\_Aiguo@iapcm.ac.cn}

\author[address2,address4,address5]{Guangcai Zhang}

\author[address1]{Yingjun Li\fnref{myfootnote2}}
\fntext[myfootnote2]{Corresponding author. Email addresses: lyj@aphy.iphy.ac.cn}

\address[address1]{State Key Laboratory for GeoMechanics and Deep Underground Engineering, China University of Mining and Technology, Beijing 100083, P.R.China}
\address[address2]{National Key Laboratory of Computational Physics, Institute of Applied Physics and Computational Mathematics, P. O. Box 8009-26, Beijing 100088, P.R.China}
\address[address3]{Center for Applied Physics and Technology, MOE Key Center for High Energy Density Physics Simulations, College of Engineering, Peking University, Beijing 100871, P.R.China}
\address[address4]{State Key Laboratory of Theoretical Physics, Institute of Theoretical
Physics, Chinese Academy of Sciences,Beijing 100190, P.R.China}
\address[address5]{State Key Laboratory of Explosion Science and Technology, Beijing
Institute of Technology, Beijing 100081, P.R.China}

\begin{abstract}
A 2-dimensional discrete Boltzmann model for combustion is presented. Mathematically, the model is composed of two coupled discrete Boltzmann equations for two species and a phenomenological equation for chemical reaction process. Physically, the model is equivalent to a reactive Navier-Stokes model supplemented by a coarse-grained model for the thermodynamic nonequilibrium behaviours. This model adopts $16$ discrete velocities. It works for both subsonic and supersonic combustion phenomena with flexible specific heat ratio. To discuss the physical accuracy of the coarse-grained model for nonequilibrium behaviours, three other discrete velocity models are used for comparisons. Numerical results are compared with analytical solutions based on both the first-order and second-order truncations of the distribution function. It is confirmed that the physical accuracy increases with the increasing moment relations needed by nonequlibrium manifestations. Furthermore, compared with the single distribution function model, this model can simulate more details of combustion.
\end{abstract}


\end{frontmatter}

%\linenumbers

\section{Introduction}

As the first controlled chemical reaction discovered by humans, combustion continues to play a dominant role in providing energy for humanity. Up to now, more than $80\%$ of the world's energy is generated from the combustion of solids, liquids, and gases \cite{Chu2012,Kuo1986}. Owing to the societal concern for energy sufficiency and environmental quality, the combustion research is extensively performed \cite{Law2006,Detonation2000,Lee2008,Lewis2012,Oran2005} in theoretical, experimental and numerical areas. Currently, combustion has become a quantitative and predictive scientific discipline. However, many fundamental problems have not yet been resolved, especially those related to various nonequilibrium behaviours \cite{Ju-Review2014,Nagnibeda2009}.

Generally, the traditional simulation of combustion is based on a hydrodynamic model combined with a phenomenological one of the chemical reaction process. The hydrodynamic model is generally a set of Euler or Navier-Stokes (NS) equations. As is known, the Euler or NS model is capable of capturing the main characteristics of hydrodynamic nonequilibrium (HNE) effects, but encounters many difficulties in describing thermodynamic nonequilibrium (TNE) effects. In the past two decades the lattice Boltzmann (LB) method has been developed as a powerful computational method for simulating multiphase flows \cite{Succi-Book,Benzi2011,Yeomans1996PRE,Yeomans1998PRL,Yeomans2013SM,Yeomans2014SM%
,Yeomans2014NP,Chikatamarla2015PRL,XuGan2015SM}, combustion phenomena \cite{Succi1997,Filippova1998,Filippova2000JCP,Filippova2000CPC,Yu2002%
,Yamamoto2002,Yamamoto2003,Yamamoto2005,Lee2006,Chiavazzo2009,Chiavazzo2010,Chiavazzo2011,ChenSheng2011,ChenSheng2012}, etc. Recently, it is being developed to investigate the trans- and supercritical fluid behaviors \cite{Rocca2015} or both the HNE and TNE simultaneously in complex systems \cite{XuGan2015SM,Review2012,ProgPhys2014,Review2015,XuYan2013,XuLin2014PRE,XuLin2014CTP,XuLin2015PRE}.\footnote{As for Ref.\cite{XuLin2015PRE}, there is a typo on the sign in front of the correction term $\hat{A_{l}}$ in Eq.(1). It should be ``$-$". The remaining derivations are correct.} Such an LB kinetic model or discrete Boltzmann model (DBM) could bring some deeper insights into nonequilibrium behaviours in various complex fluids \cite{XuGan2015SM,Review2012,ProgPhys2014,Review2015,XuYan2013,XuLin2014PRE,XuLin2014CTP,XuLin2015PRE,Succi1994}.

Previous LB models of combustion appear as a kind of new numerical scheme to solve the existing hydrodynamic models and work only for nearly incompressible systems with very low Mach number. Most of them assume that the chemical reaction does not affect the flow fields. To study the more practical cases, LB kinetic models for compressible flows \cite{Review2012,XuLin2014PRE,Novozhilov2012,XuGan2011PRE,XuChen} are needed. The first DBM for combustion and detonation was presented in 2013 \cite{XuYan2013}, which works for 2-dimensional system in Cartesian coordinates. To probe the implosion and explosion processes, a polar coordinate DBM for combustion was formulated in 2014 \cite{XuLin2014CTP}. Very recently, to make the specific heat ratio and Prandtl number flexible, we presented a multiple-relaxation-time DBM for combustion phenomena \cite{XuLin2015PRE}. Some new observations on the complex detonation processes are obtained.

However, the three existing DBMs \cite{XuYan2013,XuLin2014CTP,XuLin2015PRE} for combustion are based only on a single distribution function. Consequently, they correspond to the hydrodynamic model where only one density is used. The separate descriptions of the reactant and product depend on a process parameter $\lambda$ which is defined as the mass fraction of the product in the system. In this work we propose a double-distribution-function (DDF) DBM for combustion. Compared with the previous DBMs with single distribution function \cite{XuYan2013,XuLin2014CTP,XuLin2015PRE}, this model can be used to study more details of the combustion, such as the variations of particle number density, particle mass density, flow velocity, internal energy, temperature, and pressure of each species. Another advantage of this model is its treatment of chemical reaction. The chemical reaction rate can be a function of the densities of the reaction and products, $F(\rho^A,\rho^B)$, rather than the chemical reaction process, $F(\lambda)$. The DDF DBM has the potential to describe a more real chemical reaction.

The rest of the paper is organized as follows. In Sec. \ref{SecII} we formulate the DDF DBM based on a 2-dimensional 16-velocity discrete velocity model (D2V16 DVM). D2V13 and D2V17 DVMs are formulated for comparisons. Via the Chapman-Enskog multiscale analysis, the DDF DBM could recover reactive NS equations. The Fick's first and second laws, Stefan-Maxwell diffusion equation are also successfully confirmed by this model. Section \ref{SecIII} displays the numerical simulations. Five benchmark tests are used to validate and verify the DBM. The physical accuracy of the coarse-grained model of the TNE and the computational costs of DBM based on various DVMs are discussed in Sec. \ref{SecIV}. Section \ref{SecV} summarizes the paper.

\section{Model construction}\label{SecII}

To obtain more details of combustion procedure, we resort to DDF DBM. Recently, extensive efforts have been made to construct more accurate and stable models for multicomponent flows \cite{Shan1993,Shan1996,Shan2010,Martys1996,Sofonea2001,Luo2002,Luo2003,Xu2005PRE,Xu2005EPL,McCracken2005,Arcidiacono2006MCS,
Arcidiacono2006PRE,Arcidiacono2007,Asinari2008,Asinari2009,Li2007,Benzi2009,Kang2013,Kang2014}. In 2001 Sofonea and Sekerka \cite{Sofonea2001} proposed a nice Bhatnagar-Gross-Krook (BGK) model for isothermal binary fluid systems, where a split collision term model was discussed. In 2005 Xu \cite{Xu2005EPL} formulated a finite-difference LB method for the complete NS equations. This method is designed to simulate compressible and thermal binary fluid mixtures. However, previous DDF models can not describe compressible and thermal system with chemical reaction. In this work, we propose a DDF DBM for subsonic and supersonic combustion phenomena. This model is composed of two sets of equations. One set is two coupled discrete Boltzmann equations for thermal and high speed compressible fluid systems with two components, the other is a phenomenological equation for the chemical reaction process. Extension from a binary mixture DBM to a multispecies DBM appears straightforward \cite{Kang2013,Kang2014}.

In this section, we first introduce the fundamental relations between the physical quantities and the (discrete) distribution functions. Those relations will be employed by the DDF DBM, which is obtained as a simple coarse-grained model. The DBM uses D2V16 model constructed in the following. Meanwhile, we present two other DVMs (D2V13 and D2V17), which will be used for comparisons. Finally, the NS equations, Fick's laws, and Stefan-Maxwell diffusion equation are recovered from the kinetic model.

\subsection{Fluid system with two components}

We consider a $D$-dimensional fluid system with two components. Particle of component $\sigma$ ($=A$, $B$) has mass $m^{\sigma }$. The distribution function of component $\sigma $ reads $f^{\sigma}\equiv f^{\sigma}(\mathbf{r},\mathbf{v},\eta^2)$ at the point ($\mathbf{r}$, $\mathbf{v}$, $\eta$) in phase space. The corresponding discrete function is $f^{\sigma}_{i}\equiv f^{\sigma}_{i}(\mathbf{r},\mathbf{v}_{i},\eta^2_{i})$ with $i=1$, $2$, $\cdots $, $N$ and $N$ the total number of the discrete velocities \footnote{Because the discrete distribution function $f_i$ depends on $\eta^2_{i}$, this model gives only the right physical quantities associated with the internal energies of extra DOFs, such as the specific heat ratio $\gamma$.}. The particle number density, mass density and mean velocity of species $\sigma $ are, respectively,
\begin{equation}
n^{\sigma }=\int\int f^{\sigma }d\mathbf{v}d\eta=\sum_{i} f^{\sigma}_{i} \tt{,} \label{n_sigma}
\end{equation}
\begin{equation}
\rho ^{\sigma }=m^{\sigma }n^{\sigma }\tt{,}
\label{rho_sigma}
\end{equation}
\begin{equation}
\mathbf{u}^{\sigma }
=\frac{1}{n^{\sigma }}\int\int f^{\sigma }\mathbf{v}d\mathbf{v}d\eta
=\frac{1}{n^{\sigma }}\sum_{i} f^{\sigma}_{i}\mathbf{v}_{i}
\tt{,}
\end{equation}
where the integral is extended over all phase space ($\mathbf{v}$,$\eta $). The particle number density, mass density and hydrodynamic velocity of physical system are
\begin{equation}
n=\sum_{\sigma }n^{\sigma } \tt{,}
\end{equation}
\begin{equation}
\rho =\sum_{\sigma }\rho ^{\sigma } \tt{,}
\label{rho_0}
\end{equation}
\begin{equation}
\mathbf{u}=\frac{1}{\rho }\sum_{\sigma }\rho ^{\sigma }\mathbf{u}^{\sigma } \tt{,}
\label{u_0}
\end{equation}
respectively. The internal energy of species $\sigma $ per unit volume and the internal energy of physical system are
\begin{equation}
E^{\sigma }
=m^{\sigma }\int\int \frac{1}{2}f^{\sigma}[(\mathbf{v}-\mathbf{u})^{2}+\eta^2]d\mathbf{v}d\eta
=m^{\sigma }\sum_{i} \frac{1}{2}f^{\sigma}_{i}[(\mathbf{v}_{i}-\mathbf{u})^{2}+\eta_{i}^2]
\tt{,}
\label{Eint_sigma}
\end{equation}
\begin{equation}
E=\sum_{\sigma }E^{\sigma } \tt{,}
\end{equation}
respectively. The energy equipartition theorem gives that
\begin{equation}
E=\frac{D+I}{2}n T\tt{,}  \label{Eint_0}
\end{equation}
where $I$ is the number of extra degrees of freedom (DOFs) \footnote{In kinetic theory, the number of extra DOFs is a function of temperature, $I=I(T)$, and all DOFs are activated only at sufficiently high temperature. In this work, for simplicity, only a parameter $\eta$ is used to globally describe the internal energies in extra DOFs, and $I$ is constant. Its dependence upon temperature can be taken into account in future.}, corresponding to molecular rotation and/or vibration, whose energy level is $\eta^2/2$. Consequently, the temperature of the whole system is
\begin{equation}
T=\frac{2E}{(D+I)n}\tt{.}  \label{temperature_0}
\end{equation}
Similarly, we define
\begin{equation}
\overline{T}^{\sigma }=\frac{2E^{\sigma}}{(D+I)n^{\sigma }}
\end{equation}
as the temperature of species $\sigma$ relative to the velocity $\mathbf{u}$ per unit volume. Here the overline of $\overline{T}^{\sigma }$ is used to distinguish from the definition in Eq. (\ref{TemperatureSelf}).

The local equilibrium distribution function ${f}^{\sigma eq}$ takes the form \cite{Watari2007}
\begin{equation}
{{f}^{\sigma eq}}={{n}^{\sigma }}{{\left( \frac{{{m}^{\sigma }}}{2\pi T} \right)}^{D/2}}{{\left( \frac{{{m}^{\sigma }}}{2\pi IT} \right)}^{1/2}}\exp \left[ -\frac{{{m}^{\sigma }}{{\left( \mathbf{v}-\mathbf{u} \right)}^{2}}}{2T}-\frac{{{m}^{\sigma }}{{\eta }^{2}}}{2IT} \right]\tt{.}
\label{MostProbableDistribution}
\end{equation}
It can be proved that ${f}^{\sigma eq}$ is the most probable distribution in the system with given parameters ($m^{\sigma}$, $n^{\sigma}$, $\mathbf{u}$, $T$, $I$). Furthermore, the definitions in Eqs. (\ref{n_sigma})$-$(\ref{MostProbableDistribution}) are in line with the conservation of mass, momentum and energy.

\subsection{Discrete Boltzmann equation}

Our focus here is on the construction of a DBM suitable for combustion, but not on the chemical reaction mechanism. Since the chemical reaction is quite complex during the process of combustion. For convenience, the treatment of chemical reaction is simplified as follows:

1. The chemical reaction is irreversible and exothermic. No external force is under consideration. The electronic excitation, ionization and radiation are negligible. The combustion is temperature-dependent. The chemical reaction takes place when the temperature of chemical reactant $\overline{T}^{A}$ is larger than the ignition temperature $T_{c}$. This is a gross simplification. The ignition temperature is an artificial input parameter. Other proper model including realistic chemical kinetics can also be employed. This was done, for example, by Karlin and co-authors \cite{Chiavazzo2009,Chiavazzo2010,Chiavazzo2011}.

2. We adopt the well known Cochran's rate function \cite{Cochran1979}
\begin{equation}
\lambda'=a_{1}p^{a_{2}}(1-\lambda )+a_{3}p^{a_{4}}\lambda
(1-\lambda )  \label{Cochran}
\end{equation}
to describe the process of chemical reaction. Here the superscript $'$ is an operator denoting the material derivative $\frac{d}{dt}|_c$ on account of the effect of chemical reaction, i.e., $\lambda'=\frac{d\lambda}{dt}|_c$; And the mass fraction of product is $\lambda=\lambda^{B}=\rho^{B}/\rho$ with $\rho^{B}$ the product density; The parameters $a_{1}$, $a_{2}$, $a_{3}$ and $a_{4}$ can be adjusted to control the rate of chemical reaction.

3. The chemical reaction time is much longer than that of the kinetic process but shorter than that of the hydrodynamic flow behavior \cite{Nagnibeda2009}. Therefore, during a relatively short period of the local chemical reaction, neither mass density of physical system nor hydrodynamic velocity of the system is affected when the temperature changes with the chemical energy transformed into heat.

With the aforementioned conditions, we get the following discrete Boltzmann equation,
\begin{equation}
\frac{\partial f^{\sigma}_{i}}{\partial t}+v_{i\alpha}\frac{\partial f^{\sigma}_{i}}{\partial r_{\alpha}}=\Omega^{\sigma}_{i}+C^{\sigma}_{i}
\tt{,}
\label{DiscreteBoltzmannEquation}
\end{equation}
where the superscript $\sigma=A$ denotes the reactant and $B$ the product; $f^{\sigma}_{i}$ is the discrete nonequilibrium distribution function of component $\sigma$; $v_{i\alpha}$ ($=v_{ix}$, $v_{iy}$) the discrete velocity; $r_{\alpha}$ ($=x$, $y$) the coordinate ($D=2$). The collision term $\Omega^{\sigma}_{i}$ is given as
\begin{equation}
\Omega^{\sigma}_{i}=-\frac{1}{\tau^{\sigma}}(f^{\sigma}_{i}-f^{\sigma eq}_{i})
\tt{,}
\end{equation}
where $f^{\sigma eq}_{i}$ is the discrete equilibrium distribution function of component $\sigma$; The relaxation time is $\tau^{\sigma}=1/(n^{A}/\theta^{A}+n^{B}/\theta^{B})$, with  two relaxation parameters $\theta^{A}$ and $\theta^{B}$ to be determined \cite{Sofonea2001}. To assure local momentum conservation, the relaxation time of the two components should be equal to each other \cite{Sofonea2001}, i.e., $\tau^{\sigma}=\tau$. The chemical term is written as
\begin{equation}
C^{\sigma}_{i}=\frac{1}{\tau^{\sigma}}(f^{\sigma *eq}_{i}-f^{\sigma eq}_{i}) \tt{,}
\label{ChemicalTerm}
\end{equation}
where $f^{\sigma eq}_{i}=f^{\sigma eq}_{i}(n^{\sigma},\mathbf{u},T)$ and $f^{\sigma *eq}_{i}=f^{\sigma *eq}_{i}(n^{\sigma *},\mathbf{u},T^*)$ represent the equilibrium distribution function before and after chemical reaction, respectively. The physical quantities after reaction are
\begin{equation}
\rho^{\sigma *}=\rho^{\sigma}+\rho\lambda^{\sigma \prime}\tau^{\sigma} \tt{,}
\label{C_density}
\end{equation}
\begin{equation}
n^{\sigma *}=\frac{\rho^{\sigma *}}{m^{\sigma}} \tt{,}
\label{C_number_sigma}
\end{equation}
\begin{equation}
n^{*}=\sum_{\sigma}n^{\sigma *} \tt{,}
\label{C_number}
\end{equation}
\begin{equation}
E^{*}=E+\rho \lambda^{\prime} Q\tau^{\sigma} \tt{,}
\label{C_Energy}
\end{equation}
\begin{equation}
T^{*}=\frac{2E^{*}}{(D+I)n^{*}} \tt{,}
\label{C_Temperature}
\end{equation}
where $Q$ is the amount of heat released by the chemical reactant per unit mass; $\lambda ^{\sigma }$($=\rho ^{\sigma }/\rho $) is the mass fraction of species $\sigma $. It is clear that $\lambda^{B}=1-\lambda^{A}$ and $\lambda^{B\prime }=-\lambda^{A\prime}$.

From Eqs. (\ref{Eint_0}) and (\ref{C_Energy}), we get
\begin{equation}
E'=\rho \lambda' Q=E'_{T}+E'_{n} \tt{,}
\label{SmallE0}
\end{equation}
with
\begin{equation}
E'_{T}=\frac{D+I}{2}n'T=\frac{D+I}{2}\frac{dn}{dt}|_c T \tt{,}
\label{SmallE1}
\end{equation}
\begin{equation}
E'_{n}=\frac{D+I}{2}nT'=\frac{D+I}{2}\frac{dT}{dt}|_c n \tt{,}
\label{SmallE2}
\end{equation}
where $E'_{T}$ ($E'_{n}$) is the energy for the increase of particle number density (temperature) with fixed temperature (particle number density). When $Q=0$, we get $E'_{T}+E'_{n}=0$ and $n'/T'=-n/T$.

Equations (\ref{C_density})$-$(\ref{C_Energy}) give
\begin{equation}
n^{\prime}=\sum_{\sigma}n^{\sigma \prime}=-\sum_{\sigma}\frac{\rho\lambda^{\prime}}{m^{\sigma}} \tt{,}
\label{nprime}
\end{equation}
\begin{equation}
E^{\prime}=\rho Q \lambda^{\prime} \tt{.}
\label{Eprime}
\end{equation}
From Eqs. (\ref{temperature_0}), (\ref{nprime}) and (\ref{Eprime}), we get
\begin{equation}
T'=(\frac{2}{D+I}\frac{E}{n})'=\frac{\rho \lambda'}{n} (\frac{2Q}{D+I}-\frac{m^{A}-m^{B}}{m^{A}m^{B}}T) \tt{.}
\label{SmallT}
\end{equation}
Consequently, the temperature increases under the condition
\begin{equation}
Q>\frac{D+I}{2}\frac{m^{A}-m^{B}}{m^{A}m^{B}}T \tt{,}
\label{TemperatureIncrease}
\end{equation}
otherwise, it does not increase. When $Q=0$, the temperature reduces (increases) in the process of decomposition (combination) reaction.

In addition, the temporal derivative in discrete Boltzmann equation (\ref{DiscreteBoltzmannEquation}) is solved analytically with the first order accuracy \cite{XuLin2014PRE}, the spatial derivative in Eq. (\ref{DiscreteBoltzmannEquation}) is calculated by adopting the nonoscillatory and nonfree-parameters dissipative finite difference scheme with the second order accuracy \cite{Zhang1991}, and the chemical term (\ref{ChemicalTerm}) is calculated with the first order accuracy. Note that, at the level of the first order accuracy, the chemical term (\ref{ChemicalTerm}) is equivalent to the one proposed in our previous work \cite{XuLin2015PRE}. In fact, both this model and the previous one \cite{XuLin2015PRE} are limited to the first order accuracy. Furthermore, the method used here is different from that used in our previous works \cite{XuYan2013,XuLin2014CTP,XuLin2015PRE} to couple the chemical reaction with the flow behaviors. In our previous works \cite{XuYan2013,XuLin2014CTP,XuLin2015PRE}, $\lambda$ is obtained by solving the semiempirical evolution equation. While in this work, the result of $\lambda$ is directly given by its definition $\lambda=\rho ^{B}/\rho$. The latter approach is more reliable. Moreover, compared with the chemical term calculated in Ref. \cite{XuLin2015PRE} using $24$ discrete velocities, the chemical term given by Eq. (\ref{ChemicalTerm}) allows the DBM to employ less discrete velocities, see the following subsection. That is to say, the method presented here is more efficient.

\subsection{Discrete velocity model}\label{DVM}

Here we present three DVMs, i.e., D2V16, D2V13, and D2V17. The number of discrete velocities in the three DVMs is $N=16$, $13$, and $17$, respectively. The first DVM works for the case where the specific heat ratio $\gamma$ is adjustable, and the last two only for the case of fixed $\gamma=2$. The discrete equilibrium distribution function should satisfy the following moment relations
\begin{equation}
\mathbf{M}\ \mathbf{f}^{\sigma eq}=\mathbf{\hat{f}}^{\sigma eq} \tt{,}
\label{matrix_fcapeq}
\end{equation}
where $\mathbf{f}^{\sigma eq}=(f^{\sigma eq}_{1},f^{\sigma eq}_{2},\cdots ,f^{\sigma eq}_{N})^{T}$ is a set of discrete equilibrium distribution function; $\mathbf{\hat{f}}^{\sigma eq}=(\hat{f}^{\sigma eq}_{1},\hat{f}^{\sigma eq}_{2},\cdots , \hat{f}^{\sigma eq}_{N})^{T}$ is a set of moments provided by the discrete equilibrium distribution function; And $\mathbf{M}=(\mathbf{M}_{1},\mathbf{M}_{2},\cdots ,\mathbf{M}_{N})^{T}$ is a $N\times N$ matrix that acts as a bridge between $\mathbf{f}^{\sigma eq}$ and $\mathbf{\hat{f}}^{\sigma eq}$, with $\mathbf{M}_{i}=(m_{i1},m_{i2},\cdots ,m_{iN})$. It should be pointed out that the moment relations in Eq. (\ref{matrix_fcapeq}) are chosen to recover the hydrodynamic equations. And the number of discrete velocities in each DVM equals that of the relations in Eq. (\ref{matrix_fcapeq}). From Eq. (\ref{matrix_fcapeq}), we get the discrete equilibrium distribution functions as
\begin{equation}
\mathbf{f}^{\sigma eq}=\mathbf{M}^{-1} \mathbf{\hat{f}}^{\sigma eq} \tt{,}
\label{matrix_feq}
\end{equation}
where $\mathbf{M}^{-1}$ is the inverse matrix of $\mathbf{M}$.

\begin{figure}[tbp]
\center\includegraphics*
[bbllx=0pt,bblly=0pt,bburx=121pt,bbury=126pt,angle=0,width=0.4\textwidth]{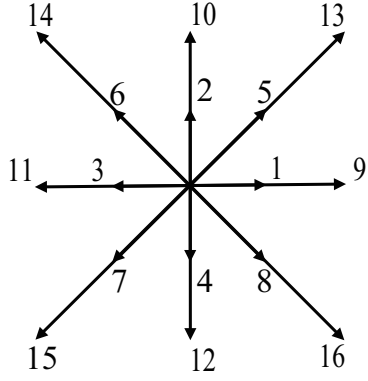}
\caption{Schematic of the D2V16 model.}
\label{Fig01}
\end{figure}

The specific elements of $\mathbf{\hat{f}}^{\sigma eq}$ and $\mathbf{M}$ in D2V16, D2V13, and D2V17 are summarized in \ref{APPENDIXA}. The discrete velocities $\mathbf{v}_{i}$ in the three DVMs are listed in Table \ref{TableI}. The elements $\eta_{i}$ in D2V16 are ($\eta_{1}$, $\eta_{2}$, $\dots$, $\eta_{N}$) $=$
($0$, $0$, $0$, $0$, $\eta_{a}$, $\eta_{a}$, $\eta_{a}$, $\eta_{a}$, $0$, $0$, $0$, $0$, $0$, $0$, $0$, $0$). The schematic of D2V16 is shown in Fig. \ref{Fig01}.
%%%%%%%%%%%%%%%%%%%%%%%%%%%%%%%%%%%%%%%%%%%%%%%%%%%%%%%%%%%%%%%%%%%%
%%%%%%%%%%%%%%%%%%%%%%%%%%%%%%%%%%%%%%%%%%%%%%%%%%%%%%%%%%%%%%%%%%%%
\begin{center}
\begin{table}[tbp]
\begin{tabular}{cc}
\hline\hline
DVM & $\left(
\begin{array}{c}
v_{1x}, v_{2x}, \dots, v_{Nx} \\
v_{1y}, v_{2y}, \dots, v_{Ny}
\end{array}
\right)$\\
\hline
D2V16 & $\left(
\begin{array}{c}
v_{a}, 0, -v_{a}, 0, v_{b}, -v_{b}, -v_{b}, v_{b}, v_{c}, 0, -v_{c}, 0, v_{d}, -v_{d}, -v_{d}, v_{d} \\
0, v_{a}, 0, -v_{a}, v_{b}, v_{b}, -v_{b}, -v_{b}, 0, v_{c}, 0, -v_{c}, v_{d}, v_{d}, -v_{d}, -v_{d}
\end{array}
\right)$\\
\hline
D2V13 & $\left(
\begin{array}{c}
v_{a}, 0, -v_{a}, 0, v_{b}, -v_{b}, -v_{b}, v_{b}, v_{c}, 0, -v_{c}, 0, v_{d} \\
0, v_{a}, 0, -v_{a}, v_{b}, v_{b}, -v_{b}, -v_{b}, 0, v_{c}, 0, -v_{c}, v_{e}
\end{array}
\right)$\\
\hline
D2V17 & $\left(
\begin{array}{c}
v_{a}, 0, -v_{a}, 0, v_{b}, -v_{b}, -v_{b}, v_{b}, v_{c}, 0, -v_{c}, 0, v_{d}, -v_{d}, -v_{d}, v_{d}, v_{e} \\
0, v_{a}, 0, -v_{a}, v_{b}, v_{b}, -v_{b}, -v_{b}, 0, v_{c}, 0, -v_{c}, v_{d}, v_{d}, -v_{d}, -v_{d}, v_{f}
\end{array}
\right)$\\
\hline\hline
\end{tabular}
\caption{Discrete velocities in D2V16, D2V13, and D2V17.}
\label{TableI}
\end{table}
\end{center}
%%%%%%%%%%%%%%%%%%%%%%%%%%%%%%%%%%%%%%%%%%%%%%%%%%%%%%%%%%%%%%%%%%%%

Remark: (I) The parameters ($v_{a}$, $v_{b}$, $v_{c}$, $v_{d}$, $\eta_{a}$) in D2V16, ($v_{a}$, $v_{b}$, $v_{c}$, $v_{d}$, $v_{e}$) in D2V13, or ($v_{a}$, $v_{b}$, $v_{c}$, $v_{d}$, $v_{e}$, $v_{f}$) in D2V17 can be adjusted to optimize the properties of the model. Their values should guarantee the existence of $\mathbf{M}^{-1}$. (II) The $16$ ($13$) relations in D2V16 (D2V13) are the necessary and sufficient conditions for the recovery of NS equations with flexible $\gamma$ (fixed $\gamma=2$), via the Chapman-Enskog expansion. While D2V17 contains another $4$ relations in addition to the $13$ ones in D2V13 needed for the recovery of NS equations with fixed $\gamma=2$. (III) Theoretically, the discrete (equilibrium) distribution function $f^{\sigma}_{i}$ ($f^{\sigma eq}_{i}$) contains more information in D2V17 than that in D2V13. Because there are more moment relations satisfied by $f^{\sigma eq}_{i}$ in D2V17 than those in D2V13. More physical quantities could be obtained from the DVM with more relations. (IV) The computational efficiency of D2V17 is lower than that of D2V13, because there are more discrete velocities in D2V17 than those in D2V13. (V) Compared with D2V33 \cite{Watari2003} and D2V65 \cite{Watari2007}, the D2V13 (or D2V17) and D2V16 can be used to recover the same NS equations with fixed and flexible specific heat ratio, respectively. But both the latter have much less discrete velocities than the former.

\subsection{Kinetic properties of DBM}

It is easy to prove that, via the Chapman-Enskog expansion, this model could recover the reactive NS equations in the hydrodynamic limit
\begin{equation}
\frac{\partial \rho ^{\sigma }}{\partial t}+\frac{\partial }{\partial
r_{\alpha }}(\rho ^{\sigma }u^{\sigma }_{\alpha })=\rho \lambda ^{\sigma\prime }\tt{,}  \label{NS_sigma_1}
\end{equation}%
\begin{eqnarray}
&&\frac{\partial }{\partial t}(\rho ^{\sigma }u^{\sigma }_{\alpha })+\frac{%
\partial }{\partial r_{\beta }}(\delta _{\alpha \beta }p^{\sigma }+\rho
^{\sigma }u^{\sigma }_{\alpha }u^{\sigma }_{\beta })+\frac{\partial }{\partial
r_{\beta }}(P^{\sigma }_{\alpha \beta }+U^{\sigma }_{\alpha \beta }+V^{\sigma }_{\alpha \beta }) \nonumber \\
&&=\rho u_{\alpha }\lambda ^{\sigma\prime }-\frac{\rho
^{\sigma }}{\tau ^{\sigma }}(u^{\sigma }_{\alpha }-u_{\alpha })\tt{,}
\label{NS_sigma_2}
\end{eqnarray}%
\begin{eqnarray}
&&\frac{\partial }{\partial t}[\rho ^{\sigma }(e^{\sigma }+\frac{1}{2}%
u^{\sigma 2})]+\frac{\partial }{\partial r_{\alpha }}[\rho ^{\sigma }u^{\sigma }_{\alpha }(e^{\sigma }+\frac{1}{2}u^{\sigma 2})+p^{\sigma }u^{\sigma }_{\alpha }]  \nonumber \\
&&-\frac{\partial }{\partial r_{\alpha }}[\kappa ^{\sigma }\frac{\partial }{%
\partial r_{\alpha }}(\frac{D+I}{2}\frac{T^{\sigma }}{m^{\sigma }}%
)-u^{\sigma }_{\beta }P^{\sigma }_{\alpha \beta }+X^{\sigma }_{\alpha}+Y^{\sigma}_{\alpha }+Z^{\sigma }_{\alpha}]  \nonumber \\
&&=\rho \lambda ^{\sigma\prime }(\frac{D+I}{2}\frac{T}{m^{\sigma }}+\frac{1%
}{2}u^{2})+\frac{D+I}{2}\frac{\rho ^{\sigma }}{m^{\sigma }}\frac{\rho
\lambda ^{\prime }}{n}(\frac{2Q}{D+I}-\frac{m^{A}-m^{B}}{m^{A}m^{B}}T) \nonumber \\
&&-\frac{\rho ^{\sigma }}{\tau ^{\sigma }}(\frac{D+I}{2}\frac{T^{\sigma }}{%
m^{\sigma }}+\frac{1}{2}u^{\sigma  2}-\frac{D+I}{2}\frac{T}{m^{\sigma }}-%
\frac{1}{2}u^{2})\tt{,}  \label{NS_sigma_3}
\end{eqnarray}
with
\begin{equation}
P^{\sigma }_{\alpha \beta }=-\mu ^{\sigma }(\frac{\partial u^{\sigma }_{\alpha }}{%
\partial r_{\beta }}+\frac{\partial u^{\sigma }_{\beta }}{\partial r_{\alpha }}-%
\frac{2\delta _{\alpha \beta }}{D+I}\frac{\partial u^{\sigma }_{\chi }}{%
\partial r_{\chi }})\tt{,}
\end{equation}%
\begin{equation}
U^{\sigma }_{\alpha \beta }=-\rho ^{\sigma }[\frac{\delta _{\alpha \beta }}{D+I%
}(u^{\sigma 2}+u^{2}-2u^{\sigma }_{\chi }u_{\chi })+u_{\alpha }u^{\sigma }_{\beta }+u^{\sigma }_{\alpha }u_{\beta }-u^{\sigma }_{\alpha }u^{\sigma }_{\beta
}-u_{\alpha }u_{\beta }]\tt{,}
\end{equation}%
\begin{equation}
V^{\sigma }_{\alpha \beta }=\frac{\lambda ^{\sigma\prime }}{\lambda
^{\sigma }}\tau ^{\sigma }U^{\sigma }_{\alpha \beta }\tt{,}
\end{equation}%
\begin{equation}
X^{\sigma }_{\alpha}=\tau ^{\sigma }(D+I+2)\frac{\rho ^{\sigma }(u^{\sigma }_{\alpha }-u_{\alpha })}{m^{\sigma }}\frac{\rho \lambda ^{\prime }%
}{n}(\frac{Q}{D+I}-\frac{m^{A}-m^{B}}{2m^{A}m^{B}}T)\tt{,}
\end{equation}%
\begin{equation}
Y^{\sigma }_{\alpha}=\frac{\rho ^{\sigma }u^{\sigma }_{\alpha }}{D+I}%
(u^{\sigma }_{\beta}-u_{\beta})^{2}+\rho ^{\sigma }(u^{\sigma }_{\alpha}-u_{\alpha })(-\frac{D+I+2}{2}\frac{T^{\sigma }-T}{m^{\sigma }}-\frac{1}{2}%
u^{\sigma 2}+\frac{1}{2}u^{2})\tt{,}
\end{equation}
\begin{equation}
Z^{\sigma}_{\alpha}=\frac{\lambda ^{\sigma\prime }}{\lambda
^{\sigma }}\tau ^{\sigma }Y^{\sigma}_{\alpha}\tt{,}
\end{equation}%
where $p^{\sigma }=n^{\sigma}T^{\sigma }$, $e^{\sigma }=(D+I)T^{\sigma }/(2m^{\sigma })$, $\mu ^{\sigma }=p^{\sigma }\tau ^{\sigma }$, and $\kappa=\gamma \mu ^{\sigma }$ are the pressure, the internal energy per unit mass, the dynamic viscosity coefficient, and heat conductivity of species $\sigma $, respectively; $\gamma =(D+I+2)/(D+I)$ is the specific heat ratio.

Performing the operator $\sum_{\sigma }$ to the two sides of Eqs. (\ref{NS_sigma_1}) $-$ (\ref{NS_sigma_3}) gives the reactive NS equations describing the whole system,
\begin{equation}
\frac{\partial \rho }{\partial t}+\frac{\partial }{\partial r_{\alpha }}%
(\rho u_{\alpha })=0\tt{,}  \label{NS_1}
\end{equation}%
\begin{equation}
\frac{\partial }{\partial t}(\rho u_{\alpha })+\frac{\partial }{\partial
r_{\beta }}\sum_{\sigma }(\delta _{\alpha \beta }p^{\sigma }+\rho ^{\sigma }u^{\sigma }_{\alpha }u^{\sigma }_{\beta })+\frac{\partial }{\partial r_{\beta }}%
\sum_{\sigma }(P^{\sigma }_{\alpha \beta }+V^{\sigma }_{\alpha \beta })=0\tt{,}
\label{NS_2}
\end{equation}%
\begin{eqnarray}
\frac{\partial }{\partial t}[\rho (e+\frac{1}{2}u^{2})]+\frac{\partial }{%
\partial r_{\alpha }}[\sum_{\sigma }{\rho ^{\sigma }u^{\sigma }_{\alpha
}(e^{\sigma }+\frac{1}{2}u^{\sigma 2})+p^{\sigma }u^{\sigma }_{\alpha }}]
\nonumber \\
-\frac{\partial }{\partial r_{\alpha }}\sum_{\sigma }[\kappa ^{\sigma }\frac{%
\partial }{\partial r_{\alpha }}(\frac{D+I}{2}\frac{T^{\sigma }}{m^{\sigma }}%
)-u^{\sigma }_{\beta }P^{\sigma }_{\alpha \beta }+Y^{\sigma }_{\alpha}+Z^{\sigma}_{\alpha}]=\rho \lambda ^{\prime }Q\tt{,}  \label{NS_3}
\end{eqnarray}%
where the internal energy of the whole system per unit mass $e=\sum_{\sigma }\rho ^{\sigma }(e^{\sigma }+u^{\sigma 2}/2)/\rho -u^{2}/2$.

For the isothermal system without chemical reaction, when the relative flow velocity between the two species is small, we have $T^{\sigma }=T$, $\lambda^{\sigma\prime }=0$, and $u^{\sigma }\approx u$. Then Eq. (\ref{NS_2}) is equivalent to
\begin{equation}
\frac{\partial }{\partial t}(\rho u_{\alpha })+\frac{\partial }{\partial
r_{\beta }}(\delta _{\alpha \beta }p+\rho u_{\alpha }u_{\beta })+\frac{%
\partial P_{\alpha \beta }}{\partial r_{\beta }}=0\tt{,}
\end{equation}
where
\begin{equation}
p=\sum_{\sigma }p^{\sigma }=nT\tt{,}
\end{equation}
\begin{equation}
P_{\alpha \beta }=\sum_{\sigma }P^{\sigma }_{\alpha \beta }=-\mu (\frac{%
\partial u_{\alpha }}{\partial r_{\beta }}+\frac{\partial u_{\beta }}{%
\partial r_{\alpha }}-\frac{2\delta _{\alpha \beta }}{D+I}\frac{\partial
u_{\chi }}{\partial r_{\chi }})\tt{,}
\end{equation}
with the dynamic viscosity coefficient of the whole system
\begin{equation}
\mu=p \tau
\label{ViscosityCoefficientSystem}
\tt{.}
\end{equation}

In addition, it is easy to demonstrate \cite{Xu2005PRE,Arcidiacono2006PRE,Glicksman2000,Williams1985} that Fick's first law
\begin{equation}
J^{\sigma }_{\alpha }=-D^{\sigma }\frac{\partial \rho ^{\sigma }}{\partial
r_{\alpha }}\tt{,}  \label{Fick1}
\end{equation}
Fick's second law
\begin{equation}
\frac{\partial \lambda ^{\sigma }}{\partial t}=D^{\sigma }\frac{\partial }{%
\partial r_{\alpha }}(\frac{\partial \lambda ^{\sigma }}{\partial r_{\alpha }%
})\tt{,}  \label{Fick2}
\end{equation}
and Stefan-Maxwell diffusion equation
\begin{equation}
M^{A}M^{B}(u^{B}_{\alpha }-u^{A}_{\alpha })=D\frac{\partial M^{A}}{\partial
r_{\alpha }}-D(\lambda ^{A}-M^{A})\frac{1}{p}\frac{\partial p}{\partial
r_{\alpha }}\tt{,}
\end{equation}%
can be derived from the DBM. Here $J^{\sigma}_{\alpha}=\rho^{\sigma} (u^{\sigma}_{\alpha}-u_{\alpha})$ is the diffusive flux of mass relative to local barycentric velocity field, $D^{\sigma }=\tau ^{\sigma }T/m^{\sigma }$ the diffusivity of species $\sigma$,  $M^{\sigma }=n^{\sigma }/n$ the mole fraction, and
\begin{equation}
D=\frac{\rho }{\rho ^{A}\rho ^{B}}M^{A}M^{B}p\tau
\label{DiffusionCoefficientSystem}
\end{equation}
the diffusion coefficient of the whole system. From Eqs. (\ref{ViscosityCoefficientSystem}) and (\ref{DiffusionCoefficientSystem}), we obtain the Schmidt number of the whole system
\begin{equation}
Sc=\frac{\mu }{\rho D}=\frac{\rho ^{A}\rho ^{B}}{M^{A}M^{B}\rho ^{2}}\tt{.}
\end{equation}%
Obviously, the Schmidt number of each species is $Sc^{\sigma }=1$ in the self-diffusion case, which is typical for gases \cite{Sofonea2001}.

Furthermore, the DBM has some intrinsic advantages over conventional NS method. For instance, (i) NS equations include nonlinear convection terms, while LB equation is in a uniformly linear form and its algorithm is easy to code. (ii) NS method often involves the solution of Poisson equation, which requires global data communication. While all the information transfer in DBM is local in time and space, so it is suitable for massively parallel computers. (iii) The DBM provides a simple method of nonequilibrium investigation, by means of calculating the velocity moments of discrete (equilibrium) distribution functions, see \ref{APPENDIXB}. In contrast, by combining the hydrodynamic equations (such as NS equations) and the evolution equations of nonequilibrium manifestations, the nonequilibrium investigation may also be made. However, it is bound to be difficult and complex, see \ref{APPENDIXB}. Although the computation costs of DVM are higher than hydrodynamic formulations, the computational overhead is minor on account of its advantages.

\section{Numerical simulations}\label{SecIII}

The validation and verification of the model are performed in this section, which is divided into five parts. The first three parts are 1-dimensional phenomena, the last two parts are 2-dimensional phenomena. Part one is an isothermal binary diffusion. Parts two and three show simulations of combustion in two different cases. Specifically, part two is for the case without reaction heat released. Part three is for a steady detonation. In the fourth part, we simulate the Kelvin-Helmholtz instability (KHI) without chemical reaction. Finally, we conduct a simulation of the Richtmyer-Meshkov instability (RMI) induced by a detonation wave. It should be mentioned that only D2V16 is used in this section.

\subsection{Binary diffusion}

Diffusion takes place when two miscible species are brought into contact. It plays an important role in combustion \cite{Lockwood1977}. The evolution of macroscopic concentration of each species obeys Fick's law. For isothermal diffusion, the following analytical solution works
\begin{equation}
M^{\sigma }=[\frac{1}{2}+\frac{\Delta M^{\sigma}}{2} erf (\frac{x}{\sqrt{4Dt}})] \tt{,}
\end{equation}
where $\Delta M^{\sigma}$ is the initial mole fraction difference and $D$ the diffusion coefficient. To compare with this solution, we simulate an isothermal diffusion here. The mixture of the two gases is initially given by the following step function,
\[
\left\{
\begin{array}{l}
(M^{A},M^{B})_{L}=(90\%,10\%) \tt{,} \\
(M^{A},M^{B})_{R}=(10\%,90\%) \tt{,}
\end{array}
\right.
\]
where the suffix $L$ indexes the left part and $R$ the right part along the axial direction $x$. It is remarked that the use of step function is generally regarded as a difficult test for the code compared to the use of smooth profiles \cite{Arcidiacono2006PRE}. The molecular masses are $m^{A}=m^{B}=1$, the relaxation parameters $\theta^{A}=\theta^{B}=10^{-3}$, and the other parameters $I=0$, ($v_{a}$, $v_{b}$, $v_{c}$, $v_{d}$, $\eta_{a}$) $=$ ($0.6$, $1.0$, $1.5$, $3.5$, $1.1$), $\Delta t=10^{-5}$, $\Delta x=\Delta y=5\times 10^{-4}$, $N_{x}\times N_{y}=80\times 1$.

%%%%%%%%%%%%%%%%%%%%%%%%%%%%%%%%%%%%%%%%%%%%%%%%%%%%%%%%%%%%%%%%%%%%
\begin{figure}[tbp]
\begin{center}
\includegraphics[bbllx=81pt,bblly=260pt,bburx=495pt,bbury=569pt,width=0.5\textwidth]{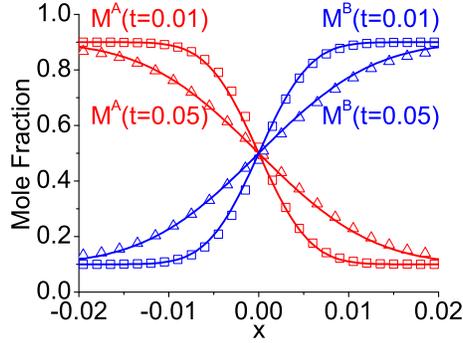}
\end{center}
\caption{Mole fractions $M^{A}$ and $M^{B}$ in the binary diffusion at two instants: $t=0.01$ and $0.05$. Symbols denote DBM simulation results and continuous lines denote the corresponding analytical solutions.}
\label{Fig02}
\end{figure}
%%%%%%%%%%%%%%%%%%%%%%%%%%%%%%%%%%%%%%%%%%%%%%%%%%%%%%%%%%%%%%%%%%%%
Figure \ref{Fig02} shows the comparison of mole fractions of the two species between the DBM simulation results and the analytical solution with $\Delta M^{\sigma}=0.8$ and $D=10^{-3}$. The simulation results at constants $t=0.01$ and $0.05$ are denoted by squares and triangles, respectively. The corresponding analytical solutions are denoted by lines. Figure \ref{Fig02} shows a satisfying agreement between the two results. Furthermore, for the same initial concentration difference, it is possible to obtain a molecular mass ratio as large as $m^{A}:m^{B}=2000:1$. It is confirmed that the DBM has a good capability of describing the interaction between two species.

\subsection{Steady shock}

Let us consider a detonation wave without chemical energy released during reaction, so it reduces to a shock wave \cite{Detonation2000,Khokhlov1999I}. Now we perform the simulation in such case. The ignition temperature is $T_{c}=1.3$. The parameters $a_{1}=1$, $a_{2}=1$, $a_{3}=200$, and $a_{4}=1$ are given here to control the chemical reaction. The initial physical quantities read
\[
\left\{
\begin{array}{l}
(\rho,u_{x},u_{y},p,\lambda)_{L}=(2.18182,1.08333,0,3.16667,1) \tt{,}  \\
(\rho,u_{x},u_{y},p,\lambda)_{R}=(1,0,0,1,0) \tt{,}
\end{array}
\right.
\]
with $L$ designating $0\leq x<0.1$ and $R$ designating $0.1\leq x<1.0$. The quantities at the two parts satisfy Hugoniot relations for shock wave. Other parameters are $I=3$, ($v_{a}$, $v_{b}$, $v_{c}$, $v_{d}$, $\eta_{a}$) $=$ ($1.9$, $1.2$, $2.5$, $3.3$, $3.3$), $\theta^{A}=10^{-3}$, $\theta^{B}=5\times10^{-4}$, $N_{x}\times N_{y}=1000\times 1$, $\Delta x=\Delta y=10^{-3}$, $\Delta t=10^{-4}$.

%%%%%%%%%%%%%%%%%%%%%%%%%%%%%%%%%%%%%%%%%%%%%%%%%%%%%%%%%%%%%%%%%%%%
\begin{figure}[tbp]
\begin{center}
\includegraphics[bbllx=2pt,bblly=3pt,bburx=142pt,bbury=329pt,width=0.5\textwidth]{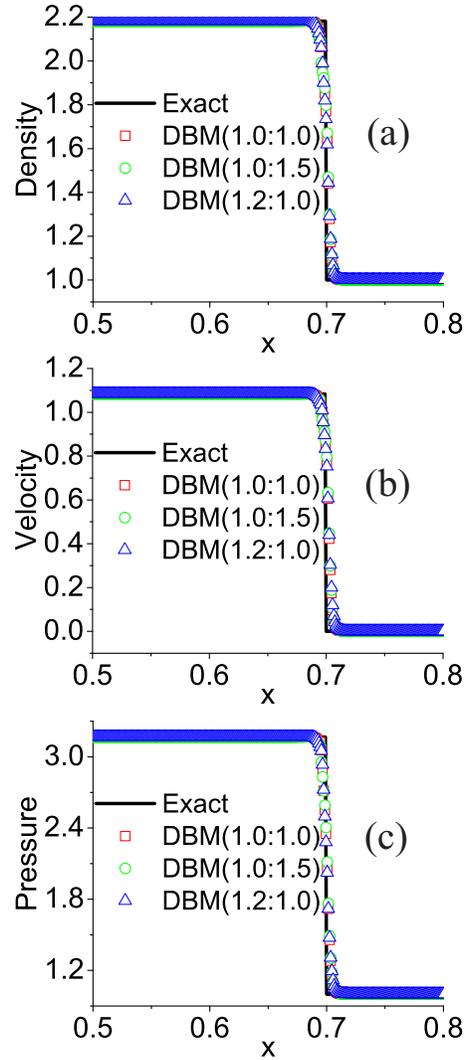}
\end{center}
\caption{Profiles of steady shock at time $t=0.3$: (a) mass density, (b) velocity $u_{x}$, (c) pressure. The analytical solution and the DBM simulation results in various cases of mass ratio $m^{A}:m^{B}$ are given in each panel. The continuous lines denote analytical solutions. The squares are for the simulation results of mass ratio $m^{A}:m^{B}=1.0:1.0$, the circles for $1.0:1.5$, and the triangles for $1.2:1.0$.}
\label{Fig03}
\end{figure}
%%%%%%%%%%%%%%%%%%%%%%%%%%%%%%%%%%%%%%%%%%%%%%%%%%%%%%%%%%%%%%%%%%%%
Figure \ref{Fig03} plots physical quantities ($\rho$, $u_{x}$, $p$) in the case of steady shock with various mass ratios at time $t=0.3$. The DBM simulation results are denoted by symbols (squares for the case $m^{A}=1.0$ and $m^{B}=1.0$, circles for $m^{A}=1.0$ and $m^{B}=1.5$, and triangles for $m^{A}=1.2$ and $m^{B}=1.0$), the analytical solution by continuous lines. It is clear to find in Fig. \ref{Fig03} that the simulation results have a satisfying agreement with the theoretical ones.

\subsection{Steady detonation}

Realistically, there is a large amount of heat released in the process of combustion or detonation in most cases, and the heat can make combustion or detonation self-sustaining \cite{Detonation2000}. As a simple example, a steady detonation is simulated here with the released reaction heat $Q=2.0$ and Mach number $2.12643$. The parameters $T_{c}=1.6$, $a_{1}=1$, $a_{2}=1$, $a_{3}=100$ and $a_{4}=1$ are given here to control the chemical reaction. The initial physical quantities read
\[
\left\{
\begin{array}{l}
(\rho,u_{x},u_{y},p,\lambda )_{L}=(1.48043,0.816497,0,3.05433,1) \tt{,}  \\
(\rho,u_{x},u_{y},p,\lambda )_{R}=(1,0,0,1,0) \tt{.}
\end{array}
\right.
\]
The quantities above satisfy Hugoniot relations for detonation wave. Other parameters are $m^{A}=1.5$, $m^{B}=1$, $I=3$, ($v_{a}$, $v_{b}$, $v_{c}$, $v_{d}$, $\eta_{a}$) $=$ ($1.9$, $1.6$, $3.2$, $6.9$, $3.9$), $\theta^{A}=2\times10^{-4}$, $\theta^{B}=10^{-4}$, $N_{x}\times N_{y}=5000\times 1$, $\Delta x=\Delta y=2\times10^{-4}$, $\Delta t=2\times10^{-5}$.

%%%%%%%%%%%%%%%%%%%%%%%%%%%%%%%%%%%%%%%%%%%%%%%%%%%%%%%%%%%%%%%%%%%
\begin{figure}[tbp]
\begin{center}
\includegraphics[bbllx=108pt,bblly=111pt,bburx=592pt,bbury=508pt,width=0.5\textwidth]%
{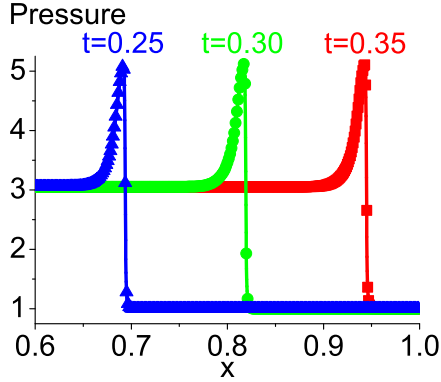}
\end{center}
\caption{The pressure along $x-$axis in the evolution of steady detonation at various instants: $t=0.25$, $0.30$, and $0.35$ from left to right, respectively.}
\label{Fig04}
\end{figure}
%%%%%%%%%%%%%%%%%%%%%%%%%%%%%%%%%%%%%%%%%%%%%%%%%%%%%%%%%%%%%%%%%%%%
Figure \ref{Fig04} gives the simulation results of pressure along $x-$axis in the evolution of the steady detonation. The results at various instants $t=0.25$, $0.30$, and $0.35$ are given from left to right, respectively. It can be found in Fig. \ref{Fig04} that the three profiles are similar to each other, which means the occurrence of a steady detonation wave. Our simulation results show that the detonation velocity is $2.512$, and the analytic solution is $2.516$. The relative error is about $0.16\%$.
%%%%%%%%%%%%%%%%%%%%%%%%%%%%%%%%%%%%%%%%%%%%%%%%%%%%%%%%%%%%%%%%%%%%
\begin{figure}[tbp]
\begin{center}
\includegraphics[bbllx=0pt,bblly=0pt,bburx=302pt,bbury=207pt,width=0.9\textwidth]{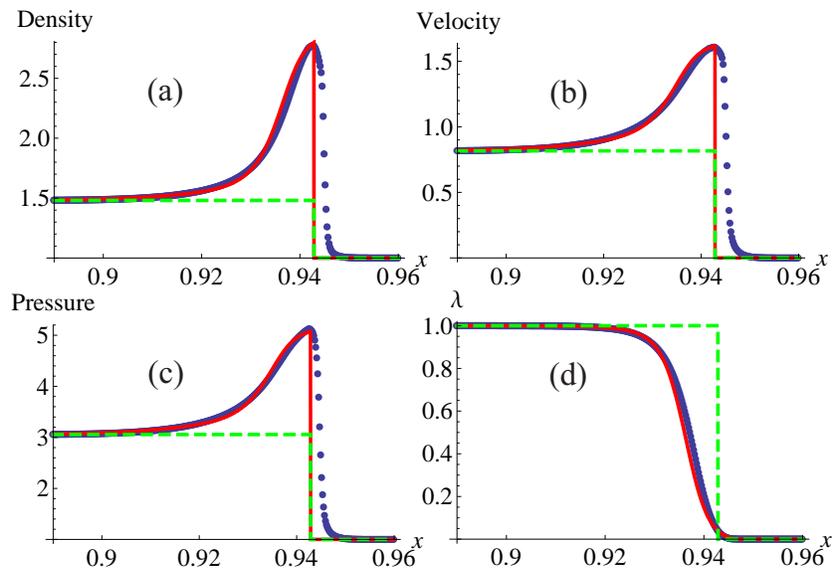}
\end{center}
\caption{Physical quantities of steady detonation at time $t=0.35$: (a) particle mass density $\rho$, (b) horizontal velocity $u_{x}$, (c) pressure $p$, (d) $\lambda $. The solid circles denote DBM simulation results, the continuous lines ZND solutions, the dashed lines CJ solutions.}
\label{Fig05}
\end{figure}
%%%%%%%%%%%%%%%%%%%%%%%%%%%%%%%%%%%%%%%%%%%%%%%%%%%%%%%%%%%%%%%%%%%%

Figure \ref{Fig05} illustrates the comparison between the simulation results and the analytical resolutions of physical quantities ($\rho$, $u_{x}$, $p$, $\lambda$) in the evolution of steady detonation at time $t=0.35$. The solid circles are for the simulation results, the continuous lines for the analytic solutions of ZND theory \cite{Detonation2000,Zeldovich1940,Neumann1942,Doering1943}, and the dashed lines for CJ theory \cite{Detonation2000,Chapman1899,Jouguet1905}. The simulation results in the detonation wave back are $\rho =1.48077$, $u_{x}=0.816470$, $p=3.05400$, $\lambda=1.0$. The deviations of them from the theoretical results are about $0.02\%$, $0.0\%$, $0.01\%$, $0.0\%$, respectively. Obviously, the simulation results have a satisfying agreement with the theoretical ones. Furthermore, it can be found that the DBM simulation results agree well with the ZND results behind the von Neumann peak. And there are a few differences between them in front of the von Neumann peak. In fact, the ZND theory ignores the viscosity and heat conduction, and it simply gives a strong discontinuity at the von Neumann peak. Therefore, it is not accurate enough. While our simulation includes the effects of viscosity, heat conduction and other kinds of relevant transportation \cite{XuLin2014CTP}. In short, the current model has the capability of simulating detonation. Further test demonstrates that the current model is successfully used to simulate detonation problem with Mach number $7.5$ (not shown here).

Moreover, compared with the previous kinetic models with single distribution function \cite{XuYan2013,XuLin2014CTP,XuLin2015PRE}, this model gives the same accuracy in predicting the ZND structure, since all of them are limited at the level of the first order accuracy. And this model can be used to study more details of the detonation structure, such as the variations of particle number density, particle mass density, flow velocity, internal energy, temperature, and pressure of each species (not shown here). Another advantage of this model is its treatment of chemical reaction. The chemical reaction rate can be connected with the densities of the reaction and products rather than a parameter $\lambda$. The DDF DBM has the capability to describe a more real chemical reaction.

\subsection{Kelvin-Helmholtz instability}

The KHI occurs when there is a relative tangential motion between two fluids with different physical parameters \cite{Chandrasekhar1968}. It plays an essential role in various situations, such as the wind blowing over the ocean, the stream structure of solar corona, the helical wave motion in ionized comet tails, the meteor entering the Earth's atmosphere, the Eagle Nebula in astrophysics, the ignition in inertial confinement fusion, the reacting mixing layers in combustion \cite{Furi2002}. Here we simulate the KHI phenomenon which is a typical 2-dimensional complex problem. The initial physical field is given as follows
\[
\left\{
\begin{array}{l}
n^{\sigma}(x)=\frac{{n^{\sigma}_{L}+n^{\sigma}_{R}}}{2}-\frac{{n^{\sigma}_{L}-n^{\sigma}_{R}}}{2}\tanh (\frac{x-x_{0}+W \cos(k y)}{{D_{\rho }}}) \tt{,} \\
\mathbf{u}(x)=\frac{{\mathbf{u}_{L}+\mathbf{u}_{R}}}{2}-\frac{\mathbf{u}_{L}{-\mathbf{u}_{R}}}{2}\tanh (\frac{x-x_{0}+W \cos(k y)}{{D_{u}}}) \tt{,} \\
p(x)=p_{L}=p_{R} \tt{,}
\end{array}
\right.
\]
where $n^{\sigma}_{L}$ ($n^{\sigma}_{R}$) is the particle number density of species $\sigma$ near the left (right) boundary; $\mathbf{u}_{L}$ and $p_{L}$ ($\mathbf{u}_{R}$ and $p_{R}$) are the velocity and pressure of the system near the left (right) boundary, respectively; $D_{\rho }$ ($D_{u}$) is the width of density (velocity) transition layer; $x_{0}$ is the averaged $x$ position of material interface; $W$ is the amplitude of initial perturbation imposed on the physical field; $k$ the perturbation wave number. The two species have the same initial velocity and temperature at the same points. Furthermore, boundary conditions for the simulation of KHI are as follows: the outflow (zero gradient) boundary conditions are adopted in the $x$ direction, the periodic boundary conditions in the $y$ direction. The parameters are chosen as $n^{A}_{L}=0.8$, $n^{A}_{R}=0.2$, $n^{B}_{L}=0.2$, $n^{B}_{R}=0.8$, $\mathbf{u}_{L}=u_{L} \mathbf{e}_{y}$, $\mathbf{u}_{R}=u_{R}\mathbf{e}_{y}$, $u_{L}=0.5$, $u_{R}=-0.5$, $D_{u}=5\Delta x$, $D_{\rho}=5\Delta x$, $W=4\Delta x$, $x_{0}=0.05$, $k=20\pi$, $\theta^{A}=10^{-5}$, $\theta^{B}=5\times 10^{-6}$, $\Delta t=5\times 10^{-6}$, $\Delta x=\Delta y=5\times 10^{-4}$, $N_{x}\times N_{y}=200\times 200$.

%%%%%%%%%%%%%%%%%%%%%%%%%%%%%%%%%%%%%%%%%%%%%%%%%%%%%%%%%%%%%%%%%%%%
\begin{figure}[tbp]
\begin{center}
\includegraphics[bbllx=0pt,bblly=0pt,bburx=280pt,bbury=291pt,width=0.8\textwidth]{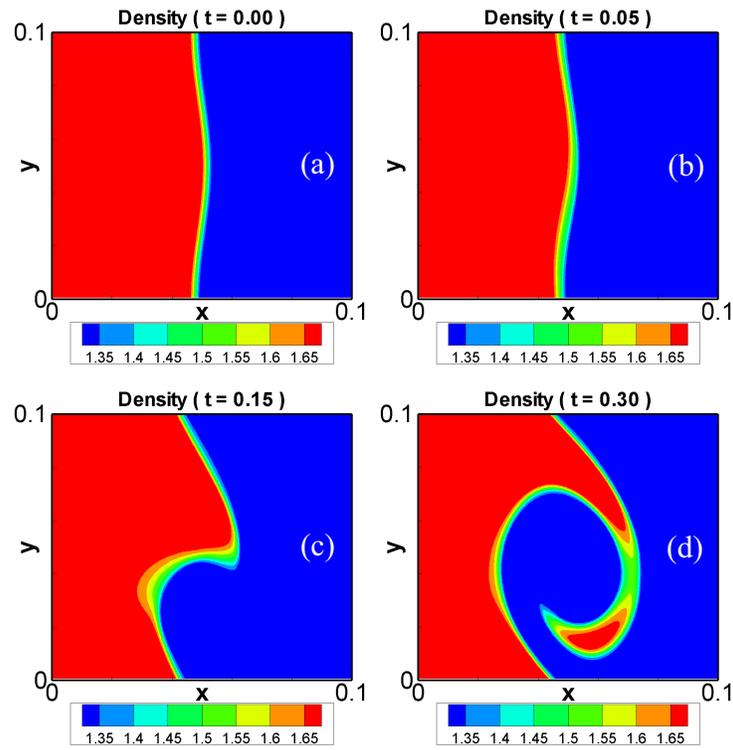}
\end{center}
\caption{Density contours in the evolution of KHI at instants $t=0.00$, $0.05$, $0.15$, and $0.30$, respectively}
\label{Fig06}
\end{figure}
%%%%%%%%%%%%%%%%%%%%%%%%%%%%%%%%%%%%%%%%%%%%%%%%%%%%%%%%%%%%%%%%%%%%
Figure \ref{Fig06} shows the density contours in the evolution of KHI at $t=0.00$, $0.05$, $0.15$, and $0.30$, respectively. Figure \ref{Fig06} (a) shows the initial density field where two fluids are divided by a single-mode sinusoidal perturbed interface. From Figs. \ref{Fig06} (a)$-$(b) we can find that the interface starts to wiggle due to the initial perturbation and the velocity shear. Figures \ref{Fig06} (b)$-$(c) show that it rolls up gradually on account of the KHI effect. It is clear to get from Figs. \ref{Fig06} (c)$-$(d) that the interfacial deformation becomes more and more significant. A larger vortex rotation is clearly observed at time $t = 0.30$. From the evolution of the continuous and smooth interface in Fig. \ref{Fig06}, it is confirmed that, the model has a good ability of capturing interface deformation.

%%%%%%%%%%%%%%%%%%%%%%%%%%%%%%%%%%%%%%%%%%%%%%%%%%%%%%%%%%%%%%%%%%%%
\begin{figure}[tbp]
\begin{center}
\includegraphics[bbllx=19pt,bblly=177pt,bburx=482pt,bbury=494pt,width=0.5\textwidth]{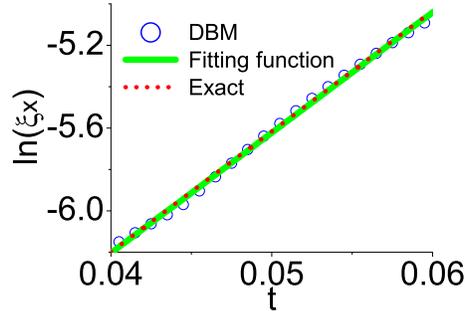}
\end{center}
\caption{The results of $\xi_x$ versus time $t$ in the linear stage of KHI.}
\label{Fig07}
\end{figure}
%%%%%%%%%%%%%%%%%%%%%%%%%%%%%%%%%%%%%%%%%%%%%%%%%%%%%%%%%%%%%%%%%%%%
To quantitatively compare the simulation results with the analytical solution, we show the logarithm of the perturbed peak kinetic energy $\xi_{x}=\rho u_{x}^2$ versus time $t$ in the  evolution of KHI, see Fig. \ref{Fig07}. The profile of $\ln(\xi_{x})$ within the linear stage ($0.04<t<0.06$) of the KHI is plotted. The circles represent the simulation results, the continuous line denotes the fitting function $F(t)=-8.53032 + 58.1616\ t$, and the dashed line is for the analytical solution $F(t)=-8.53032+2\dot{A}\ t$ where $\dot{A}=29.1392$ is half linear growth rate of $\xi_{x}$ \cite{XuGan2011PRE}. The explicit analytic formula of $\dot{A}$ is given by Eq. (18) in Ref. \cite{Wanglifeng}. It is found that the relative difference of $\dot{A}$ between the fitting function and the analytical solution is $0.20\%$, which is satisfying.

%%%%%%%%%%%%%%%%%%%%%%%%%%%%%%%%%%%%%%%%%%%%%%%%%%%%%%%%%%%%%%%%%%%%
\begin{figure}[tbp]
\begin{center}
\includegraphics[bbllx=19pt,bblly=177pt,bburx=507pt,bbury=506pt,width=0.5\textwidth]{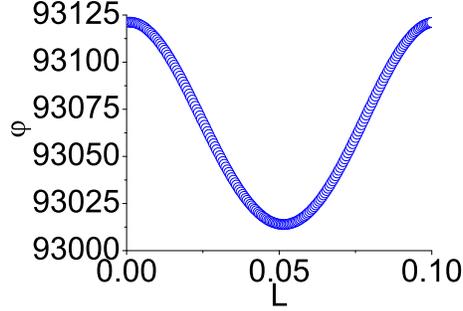}
\end{center}
\caption{The distribution of $\varphi(L,5\times 10^{-2},6\times 10^{-2})$ versus parameter $L$.}
\label{Fig08}
\end{figure}
%%%%%%%%%%%%%%%%%%%%%%%%%%%%%%%%%%%%%%%%%%%%%%%%%%%%%%%%%%%%%%%%%%%%
In order to capture the frequency $\omega$ in the evolution of KHI, we define the following correlation function
\begin{equation}
\varphi(L,t_{1},t_{2})=\int \int \rho(x,y,t_{1})\rho(x,y+L,t_{2}) dxdy
\end{equation}
where $\rho(x,y,t_{1})$ and $\rho(x,y,t_{2})$ denote the density field at times $t_{1}$ and $t_{2}$, respectively. The maximum of $\varphi(L,t_{1},t_{2})$ within $0\leq L<0.1$ is located around $\overline{u_y}(t_{2}-t_{1})$, where $\overline{u_y}$ is the average vertical velocity of the density field at time $(t_{1}+t_{2})/2$. Then the frequency is obtained
\begin{equation}
\omega=k \overline{u_y}.
\end{equation}
Figure \ref{Fig08} shows the distribution of $\varphi(L,t_{1},t_{2})$ versus $L$ with $t_{1}=5\times 10^{-2}$ and $t_{2}=6\times 10^{-2}$. The maximum of $\varphi$ is located at $L=10^{-3}$. Consequently, we obtain $\overline{u_y}=0.1$ and $\omega=6.28$ at $t=5.5\times 10^{-2}$. The analytical solution calculated by Eq. (19) in Ref. \cite{Wanglifeng} is $\omega=5.60551$. The difference between the DBM and the analytical solution mainly results from numerical errors. From $L=10^{-3}$ and $\Delta y=5\times 10^{-4}$, we find that the flow moves upward with only $L/\Delta y=2$ space steps. To reduce the numerical errors, we can decrease the space step. A further study is beyond this work.

\subsection{Richtmyer-Meshkov instability}

The RMI \cite{Richtmyer1960,Meshkov1969} occurs when a shock travels across a corrugated interface separating two fluids with different thermodynamic properties. The interplay between a shock and a flame is commonplace in many combustion systems, and the resulting instability plays a significant role in combustion \cite{Khokhlov1999II}. To simulate such a phenomenon, we give the initial
configuration as below
\[
\left\{
\begin{array}{lll}
(\rho ,u_{x},u_{y},p,\lambda )_{L} &=& (1.38837,0.57735,0,2.19162,1) \tt{,} \\
(\rho ,u_{x},u_{y},p,\lambda )_{M} &=& (1,0,0,1,0) \tt{,} \\
(\rho ,u_{x},u_{y},p,\lambda )_{R} &=& (3,0,0,1,1) \tt{,}
\end{array}
\right.
\]
where the subscripts $L$, $M$, and $R$ indicate the regions $0\leq x<0.025$,
$0.025\leq x<0.125$, and $0.125\leq x\leq 0.5$ respectively. The initial sinusoidal perturbation, $x=0.125+0.02\cos (ky)$, is applied to the physical field. The parameters $k=20\pi $, $\theta ^{A}=\theta ^{B}=2\times 10^{-5}$, $\Delta t=10^{-5}$, $\Delta x=\Delta y=10^{-3}$, $N_{x}\times N_{y}=500\times 100$. The boundary conditions are the same as those used for KHI.

%%%%%%%%%%%%%%%%%%%%%%%%%%%%%%%%%%%%%%%%%%%%%%%%%%%%%%%%%%%%%%%%%%%%
\begin{figure}[tbp]
\begin{center}
\includegraphics[bbllx=0pt,bblly=0pt,bburx=141pt,bbury=235pt,width=0.5\textwidth]{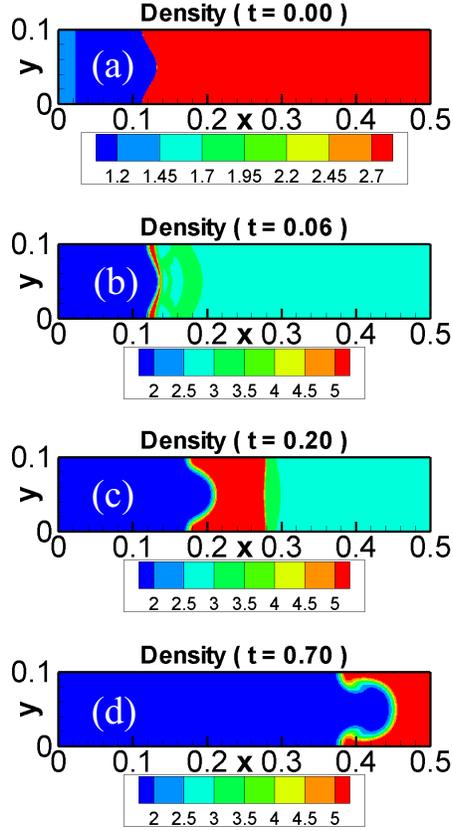}
\end{center}
\caption{Snapshots of density field in the evolution of RMI at instants $t=0.00$, $0.06$, $0.20$, and $0.70$, respectively}
\label{Fig09}
\end{figure}
%%%%%%%%%%%%%%%%%%%%%%%%%%%%%%%%%%%%%%%%%%%%%%%%%%%%%%%%%%%%%%%%%%%%
\begin{figure}[tbp]
\begin{center}
\includegraphics[bbllx=7pt,bblly=0pt,bburx=288pt,bbury=102pt,width=0.8\textwidth]{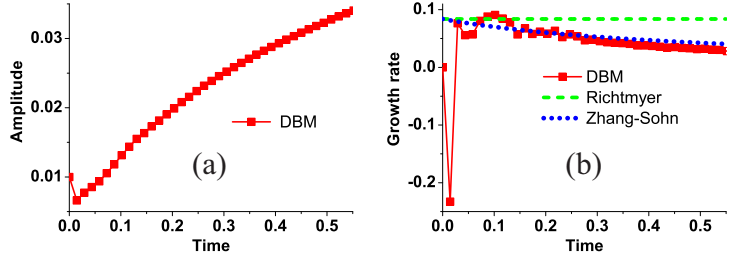}
\end{center}
\caption{Amplitude (a) and growth rate (b) of the perturbed material interface in the evolution of RMI.}
\label{Fig10}
\end{figure}
%%%%%%%%%%%%%%%%%%%%%%%%%%%%%%%%%%%%%%%%%%%%%%%%%%%%%%%%%%%%%%%%%%%%

Figure \ref{Fig09} shows the snapshots of density field in the evolution of RMI at instants $t=0.00$, $0.06$, $0.20$, and $0.70$, respectively. Figure \ref{Fig09} (a) describes the initial density field which is divided into three parts from left to right. In Fig. \ref{Fig09} (b), the detonation wave passes the perturbed material interface, and the perturbation amplitude reduces quickly due to the shock compression. After that, it starts to increase and undergo the linear and nonlinear stages, see Figs. \ref{Fig09} (b)-(d). And we can observe the occurrence of bubbles in the light medium and spikes in the heavy medium at time $t=0.70$.

To have a quantitative description of the material interface, we illustrate the evolution of amplitude and growth rate, see Fig. \ref{Fig10}. Panel (a) shows the amplitude, which is consistent with the changing density contours. Panel (b) gives the corresponding growth rate. The lines with squares denote DBM results, the dashed line denotes Richtmyer results \cite{Richtmyer1960} and the dotted line Zhang-Sohn results \cite{Zhang1997}. It can be found in panel (b) that the DBM results agree with both Richtmyer and Zhang-Sohn results in linear stage, and agree with Zhang-Sohn results in nonlinear stage. The reason is that Richtmyer model works only for linear stage, while Zhang-Sohn model works for linear and nonlinear stages.

\section{Comparison of various DVMs}\label{SecIV}

Theoretically, the more moment relations the discrete equilibrium distribution function satisfies, the more accurate the physical quantities are. However, the more discrete velocities the DVM has, the higher the computing cost is. To have a more comprehensive study on those DVMs, we need a quantitative comparison between them. It should be mentioned that this theory is independent of the value of $\gamma$. Therefore, for simplicity, we investigate D2V13 and D2V17 with $\gamma=2$ instead of D2V16 with flexible $\gamma$. Whatever the value of $\gamma$ is, the conclusion drawn from the comparison between various DVMs should be consistent.

In this section, comparisons are made between three versions of DVM, D2V13, D2V17 and D2V33. The D2V33 model was presented by Watari and Tsutahara \cite{Watari2003} in 2003. As same as D2V13 and D2V17, it only works for $\gamma =2$. This model contains five groups of particle velocities: a rest particle and four groups of octagon particles with speeds ($v_{a}<v_{b}<v_{c}<v_{d}$). The moment relations satisfied by the discrete distribution function in D2V33 contain those needed to recover NS equations. To be specific, the relations in D2V33 are exactly the same as the $13$ relations in D2V13 when the value of velocity $\mathbf{u}^{\sigma}$ is nonzero. In addition to those relations, there are other ones satisfied by the discrete distribution function in D2V33 in the case of $u_{x}^{\sigma}=0$ and/or $u_{y}^{\sigma}=0$. For example, it can be found that $\sum_{i}f^{\sigma eq}_{i}v_{iy}^{4}=\int \int f^{\sigma eq}v_{y}^{4}dv_{x}dv_{y}
$, $\sum_{i}f^{\sigma eq}_{i}v_{ix}^{3}v_{iy}=\int \int f^{\sigma eq} v_{x}^{3}v_{y}dv_{x}dv_{y}$, $\sum_{i}f^{\sigma eq}_{i}v_{ix}^{4}v_{iy}=\int
\int f^{\sigma eq}v_{x}^{4}v_{y}dv_{x}dv_{y}$, etc. in the case of $u_{y}^{\sigma}=0$.

In the following, two sets of simulations are conducted by using the three DVMs. The first set is isothermal binary diffusion, which is a nonreactive incompressible flow. The second is steady detonation, which is a compressible system with chemical reaction. The nonequilibrium manifestations are week in the former and intense in the latter. It should be mentioned that, the DBM is based on the BGK Boltzmann equation under the condition that the moments calculated from the summation of ${f}^{\sigma eq}_{i}$ are the same with those from the integration of ${f}^{\sigma eq}$. From the analytical solution $\Delta^{\sigma}_{v_{x}^2}$ in Appendix B, we can find that DBM result of $\Delta^{\sigma}_{v_{x}^2}$ is at the level of the first order accuracy. Namely, with the first order accuracy, the DBM result of $\Delta^{\sigma}_{v_{x}^2}$ is consistent with that of BGK Boltzmann equation which could reflect the deviations appearing in real gases.

\subsection{In the case of binary diffusion}

As an example, the process of binary diffusion is simulated in six tests. Test 1 and test 2 use D2V13 with adjustable parameters ($v_{a}$, $v_{b}$, $v_{c}$, $v_{d}$, $v_{e}$) $=$ ($0.3$, $1.2$, $2.3$, $0.2$, $10^{-3}$) and $1.1\times$ ($0.3$, $1.2$, $2.3$, $0.2$, $10^{-3}$), respectively; Test 3 and test 4 use D2V33 with ($v_{a}$, $v_{b}$, $v_{c}$, $v_{d}$) $=$ ($0.5$, $1.5$, $2.5$, $3.5$) and $1.1\times$ ($0.5$, $1.5$, $2.5$, $3.5$), respectively; Test 5 and test 6 use D2V17 with adjustable parameters ($v_{a}$, $v_{b}$, $v_{c}$, $v_{d}$, $v_{e}$, $v_{f}$) $=$ ($0.2$, $1.2$, $1.5$, $1.9$, $0.03$, $10^{-6}$) and $1.1\times$ ($0.2$, $1.2$, $1.5$, $1.9$, $0.03$, $10^{-6}$), respectively. Other parameters are the same as in Fig. \ref{Fig02}.

%%%%%%%%%%%%%%%%%%%%%%%%%%%%%%%%%%%%%%%%%%%%%%%%%%%%%%%%%%%%%%%%%%%%
\begin{figure}[tbp]
\begin{center}
\includegraphics[bbllx=0pt,bblly=0pt,bburx=536pt,bbury=620pt,width=0.95\textwidth]{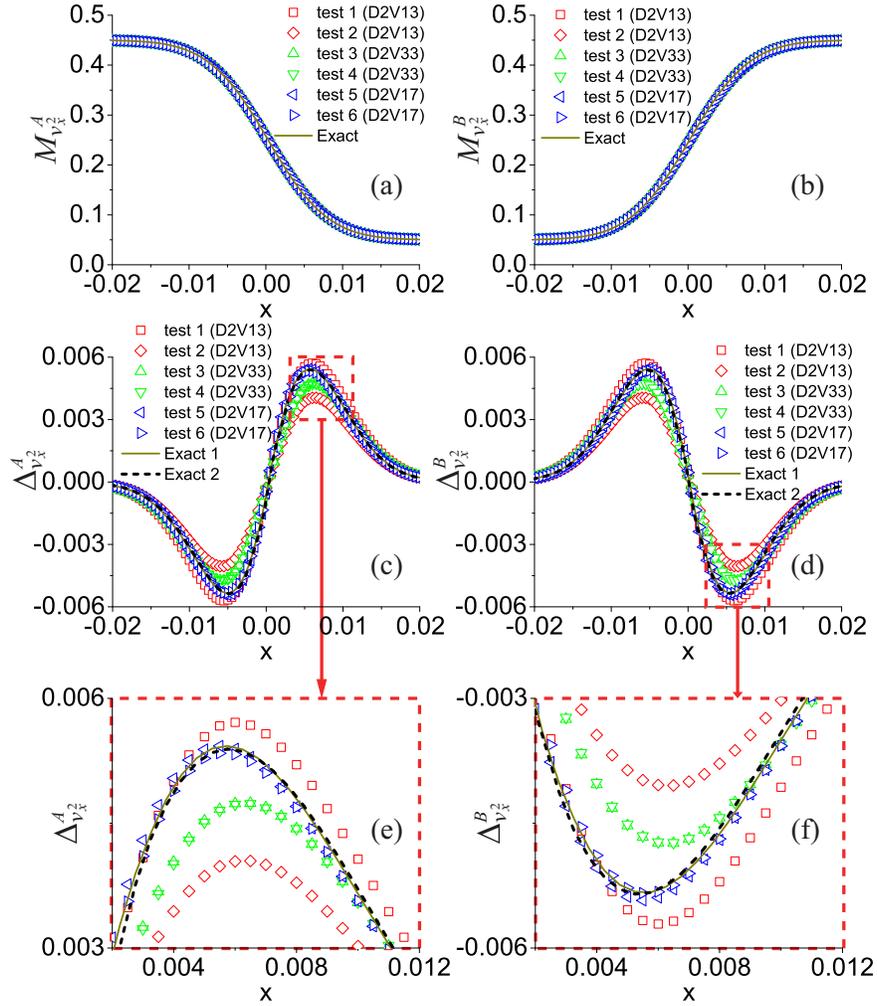}
\end{center}
\caption{Profiles of $M^{A}_{v_{x}^2}$ (a), $M^{B}_{v_{x}^2}$ (b), $\Delta^{A}_{v_{x}^2}$ (c), and $\Delta^{B}_{v_{x}^2}$ (d) in the process of binary diffusion at time $t=0.02$. Panels (e) and (f) show the enlargements of the portions in the corresponding squares in panels (c) and (d), respectively. The specific correspondences are referred to the legends.}
\label{Fig11}
\end{figure}
%%%%%%%%%%%%%%%%%%%%%%%%%%%%%%%%%%%%%%%%%%%%%%%%%%%%%%%%%%%%%%%%%%%%
Figure \ref{Fig11} shows the values of $M^{A}_{v_{x}^2}$ (a), $M^{B}_{v_{x}^2}$ (b), $\Delta^{A}_{v_{x}^2}$ (c), and $\Delta^{B}_{v_{x}^2}$ (d) in the process of binary diffusion at time $t=0.02$. Panels (e) and (f) show the enlargements of the portions in the corresponding squares in panels (c) and (d), respectively. The simulations results by using various adjustable parameters of DVMs (test 1 $-$ test 6) are plotted, with squares denoting test 1, diamonds denoting test 2, upper triangles denoting test 3, lower triangles denoting test 4, left triangles denoting test 5, right triangles denoting test 6. The continuous lines in panels (a) and (b) denote the analytical solution $M^{\sigma}_{v_{x}^2}=\frac{1}{2} (n^{\sigma} T^{\sigma}+\rho^{\sigma} u_{x}^{\sigma2})$, which is obtained by substituting $f^{\sigma}_{i}=f^{\sigma seq}_{i}$ into Eq. (\ref{Definition_Ex}). The continuous (dashed) lines in panels (c)$-$(f) denote Exact 1 (Exact 2) which is the analytical solution of $\Delta^{\sigma}_{ v_{x}^2}$  with the first (second) order accuracy. The explicit analytic formula of $\Delta^{\sigma}_{ v_{x}^2}$ is derived in \ref{APPENDIXB}. The following points should be stressed:

(I) Figures \ref{Fig11} (a)$-$(b) show that all simulation results of $M^{A}_{v_{x}^2}$ and $M^{B}_{v_{x}^2}$ coincide with their analytical solutions. While, as shown in Figs. \ref{Fig11} (c)$-$(f), the profiles of $\Delta^{A}_{v_{x}^2}$ or $\Delta^{B}_{v_{x}^2}$ depart from each other. Mathematically, the values of $M^{\sigma}_{ v_{x}^2}$ are at the level of $\mathrm{O}(\varepsilon ^{0})$, while those of $\Delta^{\sigma}_{ v_{x}^2}$ at the level of $\mathrm{O}(\varepsilon ^{1})$. Therefore, the simulation results of the former (the latter) may show slight (clear) relative difference from the analytical solutions.

(II) Figures \ref{Fig11} (c)$-$(f) show that, the simulation results of $\Delta^{A}_{ v_{x}^2}$ or $\Delta^{B}_{v_{x}^2}$ in test 5 and test 6 are closer to the analytical solutions than those in tests 1$-$4. This is because only D2V17 has all moment relations required to derive the explicit formulas of Exact 1 and Exact 2, see \ref{APPENDIXB}. The simulation results of D2V17 are the most accurate among the three DVMs. Therefore, we use the physical quantities ($n^{\sigma}$, $\mathbf{u}^{\sigma}$, $T^{\sigma}$) in test 4 to calculate the analytical solutions.

(III) It is clear in Figs. \ref{Fig11} (e)$-$(f) that test 1 and 2 have the largest relative difference and both of them are far away from the exact values; test 3 and test 4 coincide with each other and have a small departure from the exact; test 5 and 6 show few relative differences and are close to the exact. The reason is that, using the DVM with more discrete velocities, the simulation results are less dependent on the adjustable parameters. Among the three DVMs, D2V13 owns the least discrete velocities and D2V33 owns the most ones.

\subsection{In the case of steady detonation}

To further understand the nonequilibrium manifestations simulated by the DVMs, we demonstrate them in a combustion system. The initial physical quantities read
\[
\left\{
\begin{array}{l}
(\rho,u_{x},u_{y},p,\lambda )_{L}=(1.21490,0.36515,0,1.75377,1) \tt{,}  \\
(\rho,u_{x},u_{y},p,\lambda )_{R}=(1,0,0,1,0) \tt{.}
\end{array}
\right.
\]
The parameters are $Q=0.2$, $T_{c}=1.3$, $a_{1}=1$, $a_{2}=1$, $a_{3}=200$, $a_{4}=1$, $m^{A}=m^{B}=1$, $\theta^{A}=\theta^{B}=4\times10^{-4}$, $N_{x}\times N_{y}=1000\times 1$, $\Delta x=\Delta y=10^{-3}$, $\Delta t=10^{-4}$. The adjustable parameters are ($v_{a}$, $v_{b}$, $v_{c}$, $v_{d}$, $v_{e}$) $=$ ($0.7$, $1.1$, $3.9$, $3.7$, $10^{-3}$) for D2V13, ($v_{a}$, $v_{b}$, $v_{c}$, $v_{d}$) $=$ ($1$, $2$, $3$, $4$) for D2V33, ($v_{a}$, $v_{b}$, $v_{c}$, $v_{d}$, $v_{e}$, $v_{f}$) $=$ ($3.0$, $2.6$, $1.1$, $0.5$, $3.7$, $10^{-3}$) for D2V17.

%%%%%%%%%%%%%%%%%%%%%%%%%%%%%%%%%%%%%%%%%%%%%%%%%%%%%%%%%%%%%%%%%%%%
\begin{figure}[tbp]
\begin{center}
\includegraphics[bbllx=0pt,bblly=0pt,bburx=531pt,bbury=410pt,width=0.95\textwidth]{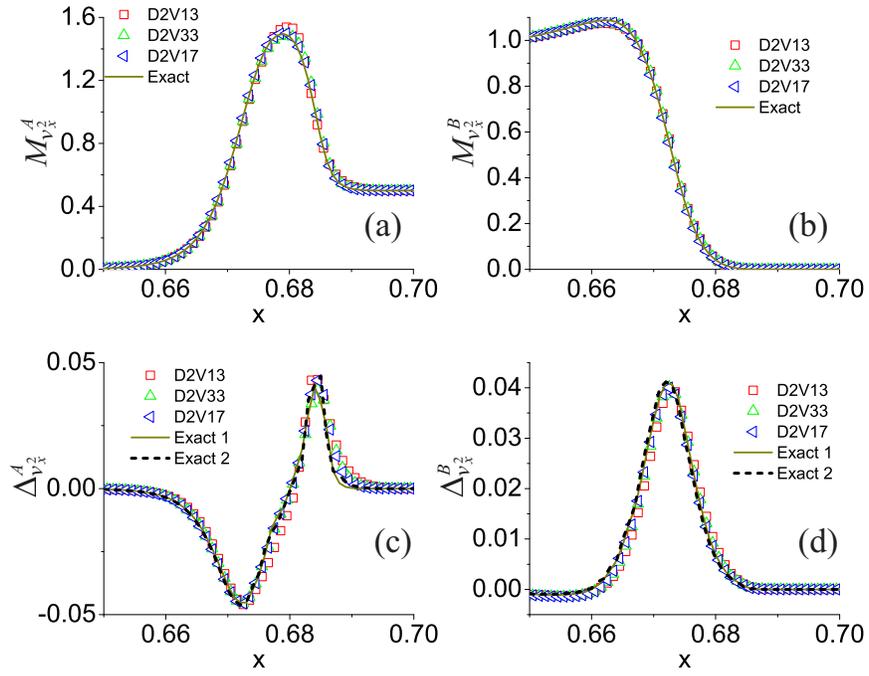}
\end{center}
\caption{Profiles of $M^{A}_{v_{x}^2}$ (a), $M^{B}_{v_{x}^2}$ (b), $\Delta^{A}_{v_{x}^2}$ (c), and $\Delta^{B}_{v_{x}^2}$ (d) in the evolution of detonation at time $t=0.3$. The specific correspondences are referred to the legends.}
\label{Fig12}
\end{figure}
%%%%%%%%%%%%%%%%%%%%%%%%%%%%%%%%%%%%%%%%%%%%%%%%%%%%%%%%%%%%%%%%%%%%
Figure \ref{Fig12} illustrates $M^{A}_{v_{x}^2}$ (a), $M^{B}_{v_{x}^2}$ (b), $\Delta^{A}_{v_{x}^2}$ (c), and $\Delta^{B}_{v_{x}^2}$ (d) in the evolution of detonation at time $t=0.3$. We can obtain the following points.

(I) The values of $\Delta^{\sigma}_{v_{x}^2}$ are at the level of $10^{-3}$ in Fig. \ref{Fig11} and at the level of $10^{-2}$ in Fig. \ref{Fig12}. Mathematically, the gradients of physical quantities ($\rho^{\sigma}$, $u^{\sigma}_{\alpha}$, $T^{\sigma}$) are much larger in the latter than in the former. Although the nonequilibrium manifestations increase by an order of magnitude, the relative difference between Exact 1 and Exact 2 in Fig. \ref{Fig12} is still small. Because this difference is proportional to $\tau^{\sigma 2}$ which is a second order small quantity, Exact 1 approaches Exact 2 when $\tau^{\sigma}$ is small enough.

(II) Figure \ref{Fig12} shows that simulation results are close to the corresponding analytical solutions. As simple coarse-grained models, all the three DVMs can be used to investigate both hydrodynamic and thermodynamic nonequilibrium behaviors simultaneously. Although the moment relations in the DVMs are limited, the accuracy of simulation results can be raised by adjusting the adjustable parameters. To obtain more accurate simulation results, we can resort to the DVM with more moment relations.

However, with discrete velocities increasing, the computational costs rise. The computing time needed for the above simulation of detonation is $7.6$ s by using D2V13, $10.9$ s by using D2V17, and $40.3$ s by using D2V33. The computational facility used here is a personal computer with Intel(R) Core(TM) $2$ CPU Q$9400$ @$2.66$GHz and RAM $4.00$ GB. It is easy to find that the computing time for D2V17 is $43\%$ more than that for D2V13, and D2V33 is about $5$ times D2V13. In addition, we further simulate the same process by using D2V16 and D2V24 model \cite{XuLin2015PRE}. The computing time is $10.6$ s for D2V16 and is $19.2$ s for D2V24. Obviously, the computational cost increases with increasing number of discrete velocities. Therefore, considering the contradiction between the operation efficiency and simulation precision, we should pursue a compromise strategy.

\section{Conclusions}\label{SecV}
A 2-dimensional kinetic model is proposed for both subsonic and supersonic combustion phenomena. Mathematically, this model is composed of two coupled discrete Boltzmann equations for fluid behaviors and a phenomenological equation for chemical reaction process. The chemical reactant is described by one distribution function, $f^{A}$, and the product by the other distribution function, $f^{B}$. The equilibrium distribution functions, $f^{Aeq}$ and $f^{Beq}$, have the same velocity and temperature.

Physically, this model is equivalent to a reactive NS model supplemented by a coarse-grained model for the thermodynamic nonequilibrium behaviours. From the kinetic model, some well-known hydrodynamic models, such as the reactive Euler and NS equations, the Fick's first and second laws, Stefan-Maxwell diffusion equation, can be easily obtained. Besides recovering hydrodynamic models, the DBM provides a simple method to measure various nonequilibrium effects. The physical accuracy of the DBM depends on the number of the kinetic moment relations of local equilibrium distribution function, instead of the number of discrete velocities. With increasing the moment relations, the DBM results in the nonequilibrium regimes become more reasonable. Additionally, it is straightforward to conduct a multiscale DBM simulation where the DVM is adaptively adjusted according to the local Knudsen number \cite{XuLin2015PRE}.

\section*{Acknowledgements}
The authors express their great appreciation to editor Prof. Thierry J. Poinsot and the anonymous reviewers for their kind and valuable suggestions, thank Profs. Zheng Chen and Cheng Wang on helpful discussions on modeling combustion, and thank Drs. Yanbiao Gan, Huilin Lai, and Zhipeng Liu for fruitful discussions on discrete Boltzmann modeling of complex flows. AX and GZ acknowledge support of the Science Foundations of National Laboratory for Science and Technology on Computational Physics, National Natural Science Foundation of China [under Grant Nos. 11475028 and 11202003], the opening project of State Key Laboratory of Explosion Science and Technology (Beijing Institute of Technology) [under Grant No. KFJJ14-1M] and the Open Project Program of State Key Laboratory of Theoretical Physics, Institute of Theoretical Physics, Chinese Academy of Sciences, China [under Grant No. Y4KF151CJ1]. YL and CL acknowledge support of National Natural Science Foundation of China [under Grant Nos. 11574390, 41472130, and 11374360], National Basic Research Program of China [under Grant No. 2013CBA01504].

\appendix

\section{}\label{APPENDIXA}

\textbf{D2V16}

The moment relations of D2V16 are in the same form of those in Ref. \cite{XuGan2013EPL}. The specific elements of $\mathbf{\hat{f}}^{\sigma eq}$ and $\mathbf{M}$ in D2V16 model are as follows,
$\hat{f}^{\sigma eq}_{1}=n^{\sigma}$,
$\hat{f}^{\sigma eq}_{2}=n^{\sigma} u_{x}$,
$\hat{f}^{\sigma eq}_{3}=n^{\sigma} u_{y}$,
$\hat{f}^{\sigma eq}_{4}=n^{\sigma} [(D+I)T/m^{\sigma}+u^{2}]$,
$\hat{f}^{\sigma eq}_{5}=n^{\sigma} (T/m^{\sigma}+u_{x}^{2})$,
$\hat{f}^{\sigma eq}_{6}=n^{\sigma} u_{x}u_{y}$,
$\hat{f}^{\sigma eq}_{7}=n^{\sigma} (T/m^{\sigma}+u_{y}^{2})$,
$\hat{f}^{\sigma eq}_{8}=n^{\sigma} u_{x}[(D+I+2)T/m^{\sigma}+u^{2}]$,
$\hat{f}^{\sigma eq}_{9}=n^{\sigma} u_{y}[(D+I+2)T/m^{\sigma}+u^{2}]$,
$\hat{f}^{\sigma eq}_{10}=3n^{\sigma} u_{x}T/m^{\sigma}+n^{\sigma} u_{x}^{3}$,
$\hat{f}^{\sigma eq}_{11}=n^{\sigma} u_{y}T/m^{\sigma}+n^{\sigma} u_{x}^{2}u_{y}$,
$\hat{f}^{\sigma eq}_{12}=n^{\sigma} u_{x}T/m^{\sigma}+n^{\sigma} u_{x}u_{y}^{2}$,
$\hat{f}^{\sigma eq}_{13}=3n^{\sigma} u_{y}T/m^{\sigma}+n^{\sigma} u_{y}^{3}$,
$\hat{f}^{\sigma eq}_{14}=n^{\sigma} [(D+I+2)T/m^{\sigma}+u^{2}]T/m^{\sigma}+n^{\sigma} u_{x}^{2}[(D+I+4)T/m^{\sigma}+u^{2}]$,
$\hat{f}^{\sigma eq}_{15}=n^{\sigma} u_{x}u_{y}[(D+I+4)T/m^{\sigma}+u^{2}]$,
$\hat{f}^{\sigma eq}_{16}=n^{\sigma} [(D+I+2)T/m^{\sigma}+u^{2}]T/m^{\sigma}+n^{\sigma} u_{y}^{2}[(D+I+4)T/m^{\sigma}+u^{2}]$;
$m_{1i}=1$,
$m_{2i}=v_{ix}$,
$m_{3i}=v_{iy}$,
$m_{4i}=v_{i}^{2}+\eta _{i}^{2}$,
$m_{5i}=v_{ix}^{2}$,
$m_{6i}=v_{ix}v_{iy}$,
$m_{7i}=v_{iy}^{2}$,
$m_{8i}=(v_{i}^{2}+\eta _{i}^{2})v_{ix}$,
$m_{9i}=(v_{i}^{2}+\eta_{i}^{2})v_{iy}$,
$m_{10i}=v_{ix}^{3}$,
$m_{11i}=v_{ix}^{2}v_{iy}$,
$m_{12i}=v_{ix}v_{iy}^{2}$,
$m_{13i}=v_{iy}^{3}$,
$m_{14i}=(v_{i}^{2}+\eta_{i}^{2})v_{ix}^{2}$,
$m_{15i}=(v_{i}^{2}+\eta _{i}^{2})v_{ix}v_{iy}$,
$m_{16i}=(v_{i}^{2}+\eta _{i}^{2})v_{iy}^{2}$.

\textbf{D2V13}

Eliminating the parameters $\eta_{i}$ and $I$ in the relations required by D2V16 model, we get $13$ linearly independent ones. The elements of $\mathbf{\hat{f}}^{\sigma eq}$ and $\mathbf{M}$ are as follows,
$\hat{f}^{\sigma eq}_{1}=n^{\sigma}$,
$\hat{f}^{\sigma eq}_{2}=n^{\sigma} u_{x}$,
$\hat{f}^{\sigma eq}_{3}=n^{\sigma} u_{y}$,
$\hat{f}^{\sigma eq}_{4}=n^{\sigma} (T/m^{\sigma}+u_{x}^{2})$,
$\hat{f}^{\sigma eq}_{5}=n^{\sigma} u_{x}u_{y}$,
$\hat{f}^{\sigma eq}_{6}=n^{\sigma} (T/m^{\sigma}+u_{y}^{2})$,
$\hat{f}^{\sigma eq}_{7}=3n^{\sigma} u_{x}T/m^{\sigma}+n^{\sigma} u_{x}^{3}$,
$\hat{f}^{\sigma eq}_{8}=n^{\sigma} u_{y}T/m^{\sigma}+n^{\sigma} u_{x}^{2}u_{y}$,
$\hat{f}^{\sigma eq}_{9}=n^{\sigma} u_{x}T/m^{\sigma}+n^{\sigma} u_{x}u_{y}^{2}$,
$\hat{f}^{\sigma eq}_{10}=3n^{\sigma} u_{y}T/m^{\sigma}+n^{\sigma} u_{y}^{3}$,
$\hat{f}^{\sigma eq}_{11}=n^{\sigma} [(D+2)T/m^{\sigma}+u^{2}]T/m^{\sigma}+n^{\sigma} u_{x}^{2}[(D+4)T/m^{\sigma}+u^{2}]$,
$\hat{f}^{\sigma eq}_{12}=n^{\sigma} u_{x}u_{y}[(D+4)T/m^{\sigma}+u^{2}]$,
$\hat{f}^{\sigma eq}_{13}=n^{\sigma} [(D+2)T/m^{\sigma}+u^{2}]T/m^{\sigma}+n^{\sigma} u_{y}^{2}[(D+4)T/m^{\sigma}+u^{2}]$;
$m_{1i}=1$,
$m_{2i}=v_{ix}$,
$m_{3i}=v_{iy}$,
$m_{4i}=v_{ix}^{2}$,
$m_{5i}=v_{ix}v_{iy}$,
$m_{6i}=v_{iy}^{2}$,
$m_{7i}=v_{ix}^{3}$,
$m_{8i}=v_{ix}^{2}v_{iy}$,
$m_{9i}=v_{ix}v_{iy}^{2}$,
$m_{10i}=v_{iy}^{3}$,
$m_{11i}=v_{i}^{2}v_{ix}^{2}$,
$m_{12i}=v_{i}^{2}v_{ix}v_{iy}$,
$m_{13i}=v_{i}^{2}v_{iy}^{2}$.

\textbf{D2V17}

Adding another $4$ relations ($\sum_{i}f^{\sigma eq}_{i}v_{ix}^{2}v_{iy}^{2}=\int \int f^{\sigma eq}v_{x}^{2}v_{y}^{2}dv_{x}dv_{y}$, $\sum_{i}f^{\sigma eq}_{i}v_{ix}v_{iy}^{3}=\int \int f^{\sigma eq}v_{x}v_{y}^{3}dv_{x}dv_{y}$, $\sum_{i}f^{\sigma eq}_{i}v_{ix}^{5}=\int \int f^{\sigma eq}v_{x}^{5}dv_{x}dv_{y}$, $\sum_{i}f^{\sigma eq}_{i}v_{iy}^{5}=\int \int f^{\sigma eq}v_{y}^{5}dv_{x}dv_{y}$) to the $13$ ones in D2V13, we obtain D2V17 model. The first $13$ elements of $\mathbf{\hat{f}}^{\sigma eq}$ and $\mathbf{M}$ in D2V17 are the same as those in D2V13, the rest $4$ ones are as follows,
$\hat{f}^{\sigma eq}_{14}=n^{\sigma} (T/m^{\sigma}+u_{x}^{2})(T/m^{\sigma}+u_{y}^{2})$,
$\hat{f}^{\sigma eq}_{15}=n^{\sigma} u_{x} u_{y} (3T/m^{\sigma}+u_{y}^{2})$,
$\hat{f}^{\sigma eq}_{16}=n^{\sigma} u_{x} (15T^{2}/m^{\sigma 2}+10u_{x}^{2})T/m^{\sigma}+u_{x}^4$,
$\hat{f}^{\sigma eq}_{17}=n^{\sigma} u_{y} (15T^{2}/m^{\sigma 2}+10u_{y}^{2})T/m^{\sigma}+u_{y}^4$;
$m_{14i}=v_{ix}^{2}v_{iy}^{2}$,
$m_{15i}=v_{ix}v_{iy}^{3}$,
$m_{16i}=v_{ix}^{5}$,
$m_{17i}=v_{iy}^{5}$.

\section{}\label{APPENDIXB}

The DBM has the ability to investigate the nonequilibrium behaviors of the physical system by using the high-order moments of $f^{\sigma}_{i}$ and
$f^{\sigma eq}_{i}$ \cite{XuGan2015SM,Review2012,ProgPhys2014,Review2015,XuYan2013,XuLin2014PRE,XuLin2014CTP,XuLin2015PRE}.
Here we introduce
\begin{equation}
M^{\sigma}_{v_{x}^2}=\frac{1}{2} \sum_{i} m^{\sigma} f^{\sigma}_{i} v_{ix}^{2}
\label{Definition_Ex} \tt{,}
\end{equation}
\begin{equation}
M^{\sigma}_{v_{y}^2}=\frac{1}{2} \sum_{i} m^{\sigma} f^{\sigma}_{i} v_{iy}^{2}
\label{Definition_Ey} \tt{,}
\end{equation}
\begin{equation}
M^{\sigma}_{\eta^2}=\frac{1}{2} \sum_{i} m^{\sigma} f^{\sigma}_{i} \eta_{i}^{2}
\label{Definition_Eeta} \tt{,}
\end{equation}
where $M^{\sigma}_{v_{x}^2}$, $M^{\sigma}_{v_{y}^2}$, $M^{\sigma}_{\eta^2}$ are the energies of species $\sigma$ in the $x$, $y$ and extra DOFs, respectively. Now let us introduce another three definitions
\begin{equation}
\Delta^{\sigma}_{v_{x}^2}
=\frac{1}{2} \sum_{i} m^{\sigma} (f^{\sigma}_{i}-f^{\sigma eq}_{i}) v_{ix}^{2}
\label{Definition_DeltaEx} \tt{,}
\end{equation}
\begin{equation}
\Delta^{\sigma}_{v_{y}^2}
=\frac{1}{2} \sum_{i} m^{\sigma} (f^{\sigma}_{i}-f^{\sigma eq}_{i}) v_{iy}^{2}
\label{Definition_DeltaEy} \tt{,}
\end{equation}
\begin{equation}
\Delta^{\sigma}_{\eta^2}=\frac{1}{2} \sum_{i} m^{\sigma}(f^{\sigma}_{i}-f^{\sigma eq}_{i})\eta^{2}
\label{Definition_DeltaEeta} \tt{,}
\end{equation}
where $\Delta^{\sigma}_{v_{x}^2}$, $\Delta^{\sigma}_{v_{y}^2}$, $\Delta^{\sigma}_{\eta^2}$ denote the departures of the energies $M^{\sigma}_{v_{x}^2}$, $M^{\sigma}_{v_{y}^2}$, $M^{\sigma}_{\eta^2}$ from the equilibrium state. Additionally, other higher order moments of $f^{\sigma}_{i}$ and ($f^{\sigma}_{i}-f^{\sigma eq}_{i}$) are referred to Refs. \cite{XuGan2015SM,XuYan2013,XuLin2014PRE,XuLin2014CTP,XuLin2015PRE}.

In fact, from the (discrete) Boltzmann equation, we can derive the relations between the nonequilibrium manifestations and physical quantities. For example, the analytical solutions $\Delta ^{\sigma }_{v_{\alpha }^{2}}$ with the first and second order accuracy are given as follows. The relations between other nonequilibrium manifestations and physical quantities can be obtained in a similar way.

Equation (\ref{DiscreteBoltzmannEquation}) gives
\begin{equation}
f^{\sigma}_{i}-f^{\sigma eq}_{i}=-\tau ^{\sigma }(\frac{\partial f^{\sigma}_{i}}{\partial t}+v_{i\alpha }\frac{\partial f^{\sigma}_{i}}{\partial
r_{\alpha }}-C^{\sigma}_{i})\mathtt{.}  \label{Deltaf}
\end{equation}%
Performing the operator $\frac{1}{2}\sum_{i}m^{\sigma }v_{i\alpha }^{2}$ to
the two sides of Eq. (\ref{Deltaf}) gives
\begin{equation}
\Delta ^{\sigma }_{v_{\alpha }^{2}}=-\frac{\tau ^{\sigma }}{2}%
\sum_{i}m^{\sigma }(\frac{\partial f^{\sigma}_{i}v_{i\alpha }^{2}}{\partial t}%
+\frac{\partial f^{\sigma}_{i}v_{i\alpha }^{2}v_{i\beta }}{\partial r_{\beta }%
}-C^{\sigma}_{i}v_{i\alpha }^{2})\mathtt{.}  \label{Delta_fvxvx}
\end{equation}%
Substituting the first-order truncation of distribution function, $f^{\sigma}_{i}\approx f^{\sigma seq}_{i}$, into the R.H.S of
Eq. (\ref{Delta_fvxvx}) gives
\begin{equation}
\Delta ^{\sigma }_{v_{\alpha }^{2}}\approx \Delta^{\sigma (1)} _{v_{\alpha }^{2}}\tt{,}
\end{equation}
\begin{equation}
\Delta^{\sigma (1)} _{v_{\alpha }^{2}}=-\frac{\tau ^{\sigma }}{2}%
\sum_{i}m^{\sigma }(\frac{\partial f^{\sigma seq}_{i}v_{i\alpha }^{2}}{%
\partial t}+\frac{\partial f^{\sigma seq}_{i}v_{i\alpha }^{2}v_{i\beta }}{%
\partial r_{\beta }}-C^{\sigma}_{i}v_{i\alpha }^{2})\mathtt{.}
\label{Delta_feqvxvx}
\end{equation}%
Here the discrete equilibrium distribution function $f^{\sigma seq}_{i}$ depends on the macroscopic parameters of species $\sigma $, i.e., $f^{\sigma seq}_{i}= f^{\sigma seq}_{i}(n^{\sigma },u^{\sigma },T^{\sigma })$.
The temperature $T^{\sigma }$ is defined as
\begin{equation}
T^{\sigma }=\frac{2E^{\sigma s}}{(D+I)n^{\sigma }} \tt{,}
\label{TemperatureSelf}
\end{equation}
\begin{equation}
E^{\sigma s}=m^{\sigma }\sum_{i}\frac{1}{2}f^{\sigma}_{i}[(\mathbf{v}_{i}-\mathbf{u}^{\sigma })^{2}+\eta_{i}^{2}] \tt{,}
\end{equation}
where $E^{\sigma s}$ is the internal energy of species $\sigma$ relative to the velocity $\mathbf{u}^{\sigma }$ per unit volume.
From Eqs. (\ref{ChemicalTerm}), (\ref{matrix_fcapeq}) and (\ref{Delta_feqvxvx}), we get the first order solution
\begin{eqnarray}
&&\Delta^{\sigma (1)} _{v_{\alpha }^{2}} =-\frac{\tau ^{\sigma }}{2}[\frac{%
\partial }{\partial t}(n^{\sigma }T^{\sigma }+\rho ^{\sigma }u_{\alpha}^{\sigma 2})  \nonumber \\
&&+\frac{\partial }{\partial r_{\beta }}(2n^{\sigma }T^{\sigma }u^{\sigma }_{\alpha }\delta _{\alpha \beta }+n^{\sigma }T^{\sigma }u^{\sigma }_{\beta }+\rho
^{\sigma }u^{\sigma 2}_{\alpha }u^{\sigma }_{\beta })]  \nonumber \\
&&+\frac{\tau ^{\sigma }}{2}[\rho \lambda ^{\sigma\prime }(\frac{T}{%
m^{\sigma }}+u_{\alpha }^{2})+\rho \lambda ^{\prime }\frac{n^{\sigma }}{n}(%
\frac{2Q}{D+I}-\frac{m^{A}-m^{B}}{m^{A}m^{B}}T)]  \tt{,}
\label{Delta_Expression}
\end{eqnarray}%
with
\begin{eqnarray}
&&\frac{\partial }{\partial t}(\rho ^{\sigma }u^{\sigma 2}_{\alpha })
=-2u^{\sigma }_{\alpha }\frac{\partial }{\partial r_{\beta }}(p^{\sigma }\delta _{\alpha \beta }+\rho ^{\sigma }u^{\sigma }_{\alpha }u^{\sigma }_{\beta
})  \nonumber \\
&&+u^{\sigma 2}_{\alpha }\frac{\partial }{\partial r_{\beta }}(\rho ^{\sigma }u^{\sigma }_{\beta })-2u^{\sigma }_{\alpha }\frac{\partial }{\partial r_{\beta }}(P^{\sigma }_{\alpha \beta }+U^{\sigma }_{\alpha \beta }+V^{\sigma }_{\alpha \beta })
\label{Rhouxux}  \nonumber \\
&&-2\frac{\rho ^{\sigma }}{\tau ^{\sigma }}u^{\sigma }_{\alpha }(u^{\sigma }_{\alpha }-u_{\alpha })-\rho u^{\sigma 2}_{\alpha }\lambda ^{\sigma\prime
}+2\rho u_{\alpha }u^{\sigma }_{\alpha }\lambda ^{\sigma\prime }
\tt{,}
\end{eqnarray}%
\begin{eqnarray}
&&\frac{\partial }{\partial t}(n^{\sigma }T^{\sigma })=\frac{\partial }{%
\partial r_{\alpha }}[\frac{\kappa ^{\sigma }}{m^{\sigma }}\frac{\partial
T^{\sigma }}{\partial r_{\alpha }}+\frac{2}{D+I}(-u^{\sigma }_{\beta }P^{\sigma }_{\alpha \beta }+X^{\sigma }_{\alpha \beta }  \nonumber \\
&&+Y^{\sigma }_{\alpha \beta
}+Z^{\sigma }_{\alpha \beta })]-\frac{\partial }{\partial r_{\alpha }}{\rho ^{\sigma }u^{\sigma }_{\alpha }(%
\frac{T^{\sigma }}{m^{\sigma }}+\frac{u^{\sigma 2}}{D+I})+\frac{2}{D+I}%
p^{\sigma }u^{\sigma }_{\alpha }}  \nonumber \\
&&+\rho \lambda ^{\sigma\prime }(\frac{T}{m^{\sigma }}+\frac{u^{2}}{D+I}%
)+\rho \lambda ^{\prime }\frac{n^{\sigma }}{n}(\frac{2Q}{D+I}-\frac{%
m^{A}-m^{B}}{m^{A}m^{B}}T)  \nonumber \\
&&-\frac{\rho ^{\sigma }}{\tau ^{\sigma }}(\frac{T^{\sigma }-T}{m^{\sigma }}+%
\frac{u^{\sigma 2}-u^{2}}{D+I})+\frac{2u^{\sigma }_{\alpha }}{D+I}\frac{%
\partial }{\partial r_{\beta }}(p^{\sigma }\delta _{\alpha \beta }+\rho
^{\sigma }u^{\sigma }_{\alpha }u^{\sigma }_{\beta })  \nonumber \\
&&-\frac{u^{\sigma 2}_{\alpha }}{D+I}\frac{\partial }{\partial r_{\beta }}%
(\rho ^{\sigma }u^{\sigma }_{\beta })+\frac{2u^{\sigma }_{\alpha }}{D+I}\frac{%
\partial }{\partial r_{\beta }}(P^{\sigma }_{\alpha \beta }+U^{\sigma }_{\alpha
\beta }+V^{\sigma }_{\alpha \beta })  \nonumber \\
&&+\frac{\rho u^{\sigma 2}_{\alpha }\lambda ^{\sigma\prime }}{D+I}-\frac{%
2\rho u_{\alpha }u^{\sigma }_{\alpha }\lambda ^{\sigma\prime }}{D+I}+\frac{%
2\rho ^{\sigma }u^{\sigma }_{\alpha }}{D+I}\frac{1}{\tau ^{\sigma }}(u^{\sigma }_{\alpha }-u_{\alpha })\tt{,}  \label{nT}
\end{eqnarray}%
which can be obtained from Eqs. (\ref{NS_sigma_1}) $-$ (\ref{NS_sigma_3}).

In addition, Eq. (\ref{DiscreteBoltzmannEquation}) gives the second-order truncation of distribution function,
\begin{equation}
f^{\sigma}_{i}=f^{\sigma seq}_{i}-\tau ^{\sigma }(\frac{\partial f^{\sigma eq}_{i}}{\partial t}+v_{i\alpha }\frac{\partial f^{\sigma eq}_{i}}{%
\partial r_{\alpha }}+S^{\sigma}_{i}-C^{\sigma}_{i})\tt{,}
\label{CE_DBM}
\end{equation}%
\begin{equation}
S^{\sigma}_{i}=\frac{1}{\tau ^{\sigma }}(f^{\sigma seq}_{i}-f^{\sigma eq}_{i})\tt{.}  \label{CouplingTerm}
\end{equation}
Substituting Eq. (\ref{CE_DBM}) into the R.S.H of Eq. (\ref{Delta_fvxvx}) gives the second order solution,
\begin{equation}
\Delta ^{\sigma }_{v_{\alpha }^{2}}\approx \Delta^{\sigma (1)} _{v_{\alpha }^{2}}+\Delta^{\sigma (2)} _{v_{\alpha }^{2}}\tt{,}
\end{equation}%
\begin{eqnarray}
&&\Delta^{\sigma (2)} _{v_{\alpha }^{2}}=-{{\tau }^{\sigma }}%
\frac{\partial }{\partial t}[\Delta ^{\sigma (1)}_{v_{\alpha }^{2}}
-\frac{1}{2}(n^{\sigma}T^{\sigma}-n^{\sigma}T+\rho^{\sigma}u^{\sigma2}_{\alpha}
-\rho^{\sigma}u^{2}_{\alpha})] \nonumber \\
&&+\frac{\tau ^{\sigma }}{2}%
\frac{\partial }{\partial r_{\beta }}\sum_{i}\tau ^{\sigma }m^{\sigma }(%
\frac{\partial f^{\sigma seq}_{i}v_{i\alpha }^{2}v_{i\beta }}{\partial t} \nonumber \\
&&+\frac{\partial f^{\sigma seq}_{i}v_{i\alpha }^{2}v_{i\beta }v_{i\chi }}{%
\partial r_{\chi }}+S^{\sigma}_{i}v_{i\alpha }^{2}v_{i\beta }-C^{\sigma}_{i}v_{i\alpha }^{2}v_{i\beta })\mathtt{.}  \label{Delta2_temp1}
\end{eqnarray}
From Eqs. (\ref{ChemicalTerm}), (\ref{matrix_fcapeq}) and (\ref{CouplingTerm}), we get
\begin{eqnarray}
&&\sum_{i}S^{\sigma}_{i}v_{i\alpha }^{2}v_{i\beta }=\frac{1}{\tau ^{\sigma }}%
[n^{\sigma }(2u^{\sigma }_{\alpha }\delta _{\alpha \beta }+u^{\sigma }_{\beta })%
\frac{T^{\sigma }}{m^{\sigma }}+n^{\sigma }u^{\sigma 2}_{\alpha }u^{\sigma }_{\beta }]  \nonumber \\
&&-\frac{1}{\tau ^{\sigma }}[n^{\sigma }(2u_{\alpha }\delta _{\alpha \beta
}+u_{\beta })\frac{T}{m^{\sigma }}+n^{\sigma }u_{\alpha }^{2}u_{\beta }]
\tt{,}
\label{B_vx**2*vy}
\end{eqnarray}%
\begin{eqnarray}
&&\sum_{i}C^{\sigma}_{i}v_{i\alpha }^{2}v_{i\beta }=\frac{\rho \lambda
^{\prime }}{m^{\sigma }}\frac{n^{\sigma }}{n}(2u_{\alpha }\delta _{\alpha
\beta }+u_{\beta })(\frac{2Q}{D+I}-\frac{m^{A}-m^{B}}{m^{A}m^{B}}T)
\nonumber \\
&&+\frac{\rho \lambda ^{\sigma\prime }}{m^{\sigma }}(2u_{\alpha }\frac{T%
}{m^{\sigma }}\delta _{\alpha \beta }+u_{\beta }\frac{T}{m^{\sigma }}%
+u_{\alpha }^{2}u_{\beta })
\tt{.}
\label{C_vx**2*vy}
\end{eqnarray}

For $1$-dimensional problem, Eq. (\ref{Delta2_temp1}) reduces to
\begin{eqnarray}
&&\Delta^{\sigma (2)} _{v_{\alpha }^{2}}=-{{\tau }^{\sigma }}%
\frac{\partial }{\partial t}[\Delta ^{\sigma (1)}_{v_{\alpha }^{2}}
-\frac{1}{2}(n^{\sigma}T^{\sigma}-n^{\sigma}T+\rho^{\sigma}u^{\sigma2}_{\alpha}
-\rho^{\sigma}u^{2}_{\alpha})] \nonumber \\
&&+\frac{\tau ^{\sigma }}{2}%
\frac{\partial }{\partial r_{\alpha }}\sum_{i}\tau ^{\sigma }m^{\sigma }(%
\frac{\partial f^{\sigma seq}_{i}v_{i\alpha }^{3}}{\partial t}
+\frac{\partial f^{\sigma seq}_{i}v_{i\alpha }^{4}}{\partial r_{\alpha }}%
+S^{\sigma}_{i}v_{i\alpha }^{3}-C^{\sigma}_{i}v_{i\alpha }^{3})\mathtt{.}
\label{Delta2_temp2}
\end{eqnarray}%
Let us suppose that $f^{\sigma seq}_{i}$ satisfies the following relation,
\begin{equation}
\int \int f^{\sigma seq}v_{\alpha }^{4}d\mathbf{v}d\eta =3n^{\sigma }%
\frac{T^{\sigma 2}}{m^{\sigma 2}}+6n^{\sigma }u^{\sigma 2}_{\alpha }%
\frac{T^{\sigma }}{m^{\sigma }}+n^{\sigma }u^{\sigma 4}_{\alpha}=\sum_{i}f^{\sigma seq}_{i}v_{i\alpha }^{4}\tt{.}
\label{moment_vx**4}
\end{equation}%
From Eqs. (\ref{matrix_fcapeq}) and (\ref{B_vx**2*vy}) $-$ (\ref{moment_vx**4}), we get
\begin{eqnarray}
&&\Delta ^{\sigma(2)}_{v_{\alpha }^{2}}=-{{\tau }^{\sigma }}%
\frac{\partial }{\partial t}[\Delta ^{\sigma (1)}_{v_{\alpha }^{2}}
-\frac{1}{2}(n^{\sigma}T^{\sigma}-n^{\sigma}T+\rho^{\sigma}u^{\sigma2}_{\alpha}
-\rho^{\sigma}u^{2}_{\alpha})] \nonumber \\
&&+\frac{{{\tau }^{\sigma }}}{2}\frac{\partial }{\partial r_{\alpha }}{{\tau
}^{\sigma }[}\frac{\partial }{\partial t}\left( 3{{n}^{\sigma }}{{u}^{\sigma }_{\alpha }}{{T}^{\sigma }}+{{\rho }^{\sigma }}u^{\sigma 3}_{\alpha } \right) \nonumber \\
&&+\frac{\partial }{\partial r_{\alpha }}(3\rho ^{\sigma }\frac{T^{\sigma 2}}{m^{\sigma 2}}+6n^{\sigma }u^{\sigma 2}_{\alpha }T^{\sigma }+\rho
^{\sigma }u^{\sigma 4}_{\alpha })  \nonumber \\
&&+\frac{1}{{{\tau }^{\sigma }}}\left( 3{{n}^{\sigma }}{{u}^{\sigma }_{\alpha }}%
{{T}^{\sigma }}+{{\rho }^{\sigma }}u^{\sigma 3}_{\alpha }-3{{n}^{\sigma }}{{u%
}_{\alpha }}T-{{\rho }^{\sigma }}u_{\alpha }^{3}\right)   \nonumber \\
&&-3\rho {\lambda }^{\prime }{{u}_{\alpha }}\frac{{{n}^{\sigma }}}{n}(\frac{%
2Q}{D+I}-\frac{{{m}^{A}}-{{m}^{B}}}{{{m}^{A}}{{m}^{B}}}T)-\rho {{{{\lambda }%
^{\sigma\prime }}}(3{{u}_{\alpha }}\frac{T}{{{m}^{\sigma }}}+u_{\alpha
}^{3})}]
\tt{.}
\end{eqnarray}

Mathematically, $\Delta^{\sigma (1)} _{v_{\alpha }^{2}}$ is proportional to $\tau ^{\sigma }$ at the level of $\mathrm{O}(\varepsilon ^{1})$, and $\Delta^{\sigma (2)} _{v_{\alpha }^{2}}$ is proportional to $\tau^{\sigma 2}$ at the level of $\mathrm{O}(\varepsilon ^{2})$.

It should be pointed that the moment relation in Eq. (\ref{moment_vx**4}) is satisfied by the discrete distribution function in D2V17, but not in D2V13 or D2V33. Performing the operator $\sum_{i}m^{\sigma }v_{i\alpha }^{4}$
to the expression of $f^{\sigma seq}_{i}$ in D2V33 \cite{Watari2003} gives
\begin{equation}
\sum_{i}f^{\sigma seq}_{i}v_{i\alpha }^{4}=3n^{\sigma }\frac{T^{\sigma 2}%
}{m^{\sigma 2}}+6n^{\sigma }u^{\sigma 2}_{\alpha }\frac{T^{\sigma }}{%
m^{\sigma }}+\frac{9}{8}n^{\sigma }u^{\sigma 4}_{\alpha }\mathtt{.}
\end{equation}%
The difference between the above equation and Eq. (\ref{moment_vx**4}) is $\frac{1}{8}n^{\sigma }u^{\sigma 4}_{\alpha }$. If $u^{\sigma }_{\alpha }$ is small enough, the difference is negligible, otherwise not negligible.

Note that the more moment relations the DVM has, the more accurate the nonequilibrium manifestations are. One can simply increase the physical accuracy of the DBM in describing TNE via using more kinetic moment relations. A further discussion is beyond this work.

%\section*{References}

\end{document}